\documentstyle[aps,prb,epsf,floats,twocolumn,eqsecnum]{revtex}

\begin{document}
\draft
\title{The random magnetic flux problem in a quantum wire}

\author{Christopher Mudry, P. W. Brouwer}
\address{Lyman Laboratory of Physics, Harvard University, Cambridge MA 02138}
\author{ Akira Furusaki}
\address{Yukawa Institute for Theoretical Physics, Kyoto University,
Kyoto 606-8502, Japan\cite{address}\\
 and Department of Physics, Stanford University, Stanford,
CA 94305\\
{\rm November 16, 1998}
\medskip \\ \parbox{14cm}{\rm
The random magnetic flux problem on a lattice and
in a quasi one-dimensional (wire) geometry
is studied both analytically and numerically.
The first two moments of the conductance are obtained analytically.
Numerical simulations for the average and variance of the conductance agree
with the theory. 
We find that the center of the band $\varepsilon=0$ plays a special role.
Away from $\varepsilon=0$, transport properties are those of a disordered
quantum wire in the standard unitary symmetry class.
At the band center $\varepsilon=0$, 
the dependence on the wire length of the conductance departs
from the standard unitary symmetry class and is governed by a new universality
class, the chiral unitary symmetry class. The most remarkable property
of this new universality class is the existence of an even-odd effect
in the localized regime:
Exponential decay of the average conductance for an even number of channels 
is replaced by algebraic decay for an odd number of channels. \smallskip }}
\narrowtext
\maketitle 

\section{Introduction}
\label{sec:Introduction}

The concepts of scaling \cite{Edwards72,Licciardello75,Wegner76} and of
the renormalization group\cite{Abrahams79} have provided crucial
insights into the localization properties of a quantum particle in a
random but static environment.\cite{Lee85} 
Beyond a typical length scale depending on the microscopic details of the
disorder, the localization problem can be described by an effective
field theory that is uniquely specified by the dimensionality of space
and the fundamental symmetries of the microscopic
Hamiltonian.\cite{Wegner79} Correspondingly, the disorder is said to
belong to the orthogonal, unitary and symplectic ensembles, depending
on whether time reversal symmetry and spin-orbit coupling are present
or not.\cite{Wegner79,Hikami80,Dyson62}

However, not all disordered systems belong to one of these three
standard symmetry classes. One example is the Integer Quantum Hall
Effect, for which the scaling theory in the unitary universality class
cannot explain the observed jumps in the Hall
resistance,\cite{Huckestein95} since it predicts that all states are
localized in two-dimensions.  Instead, a new scaling theory was
proposed for the Integer Quantum Hall Effect, where, in addition to the
longitudinal conductivity that controls the scaling flow in the unitary
ensemble, the Hall conductivity appears as a second
parameter.\cite{Khmelnitskii83,Levine83}

In this paper we consider a different example. It is the so-called
random flux model, which describes the localization properties of a
particle moving in a plane perpendicular to a static magnetic field of
random amplitude and vanishing
mean.\cite{Lee81,Pryor92,Altshuler92,Khveshchenko93,Gavazzi93,Ohtsuki93,Sugiyama93,DKKLee94,Aronov94,YBKim95,Verges96,Yakubo96,Batsch98,Avishai93,Kalmeyer93,Zhang94,Kawarabayashi95,Liu95,Sheng95,KunYang97,Xie98,Miller96,Furusaki98}
In the literature, different points of view have been offered with
regard to the localization properties and the appropriate symmetry class of the
random flux problem. In Refs.\ \onlinecite{Sugiyama93,DKKLee94,Aronov94,YBKim95,Verges96,Yakubo96,Batsch98}
it has been claimed that, since the magnetic field has a vanishing mean,
the only effect of the random magnetic field is to break time reversal
invariance, and hence that the localization properties are those of
the standard unitary symmetry class.
On the other hand, Zhang and Arovas\cite{Zhang94} have argued that this
argument might be too naive and that a scaling theory closely related
to that of the Kosterlitz-Thouless transition controls the localization
properties of the random magnetic flux problem.  They predicted
that states are localized in the tails of the spectrum
whereas close to the center of the band a line of critical
points of the Kosterlitz-Thouless type is formed. Related point of views can be
found in Refs.\
\onlinecite{Avishai93,Kalmeyer93,Zhang94,Kawarabayashi95,Liu95,Sheng95,KunYang97,Xie98}. Finally, it has been proposed in Ref.\ \onlinecite{Miller96}
that the random flux model shows critical behavior at the band center $\varepsilon=0$ only,
whereas its localization properties are those of the unitary
ensemble for energies $\varepsilon \neq 0$.

In the third scenario, the behavior at $\varepsilon=0$ is governed by
an additional symmetry, the so-called chiral or particle-hole
symmetry.  The chiral symmetry can also be found in the related problem of
a particle hopping on a lattice with random (real)
hopping amplitudes.\cite{Eilmes98}
In the one-dimensional version of this problem, it is well established
that the ensemble-averaged density of states diverges at the band
center $\varepsilon=0$\cite{Theo76,Eggarter78} and that
the ensemble averaged conductance decays algebraically with the length
$L$ of the system.\cite{Mathur97} For comparison, 
in the unitary symmetry class, the
density of states is continuous at $\varepsilon=0$,\cite{Wegner81} 
while the conductance decays exponentially with $L$. 
(The one-dimensional random-hopping problem has been studied in many 
incarnations, cf. Refs.\ 
\onlinecite{Dyson53,McCoyWu68,Smith70,Shankar87,Fisher94,McKenzie96,Balents97}.) 
For two-dimensional systems, the effect of the  chiral
symmetry was studied by Gade and Wegner\cite{Gade93} 
(see also Refs.\ 
\onlinecite{Hikami93,Minakuchi96,Ludwig94,Nersesyan94,Mudry96,Chamon96,Kogan96,Castillo97,Hatsugai97,Caux98,Mudry98}). They argued that the presence
of the chiral symmetry results in three new symmetry classes, called
chiral orthogonal, chiral unitary, and chiral symplectic.
For disordered systems with chiral unitary symmetry, 
all states are localized except
at the singular energy $\varepsilon=0$ at which the average 
density of states diverges.
The relevance of the chiral unitary symmetry class to the random flux 
problem was pointed out by Miller and Wang.\cite{Miller96} 
(Only the chiral unitary class is of relevance, since
time-reversal symmetry is broken in the random flux model.)

For the two-dimensional random-flux problem, sufficiently accurate
numerical data are notoriously hard to obtain.  Although a consensus
has emerged that states are localized in the tails of the spectrum, it
is impossible to decide solely on the basis of numerical simulations
whether states are truly delocalized upon approaching the center of the
band, or only deceptively appear so as the localization length is much
larger than the system sizes that are accessible to the current
computers.  Moreover, it is very easy to smear out a diverging
density of states in a
numerical simulation (compare Refs. \onlinecite{Pryor92,Verges96} and
\onlinecite{Furusaki98}). In short, no conclusion has been reached in
the debate about the localization properties of the two-dimensional
random flux problem.

Here, we focus on the simpler problem of the random flux problem on a
lattice and in a quasi one-dimensional geometry of a (thick) quantum wire 
with weak disorder, and restrict our attention to
transport properties, notably the conductance $g$.
For a wire geometry, numerical simulations can be performed with very high
accuracy, and very good statistics can be obtained. Moreover, precise
theoretical predictions for the transport properties can be made, both
for the unitary symmetry class, and for the chiral unitary symmetry
class. The wire geometry allows us to quantitatively compare the
analytical predictions for the various symmetry classes and the
numerical simulations for the random flux model.  This comparison shows that, 
away from the critical energy $\varepsilon=0$, the $L$-dependence
of the average and variance of the conductance $g$ are those of the 
unitary ensemble.  At the band center $\varepsilon=0$, 
$\langle g\rangle$ and $\mbox{var}\, g$ 
are given by the chiral unitary ensemble. 
Hence, we unambiguously show that in a quasi one-dimensional geometry, 
the localization properties of the random flux model 
are described by the
third scenario above, in which the $\varepsilon=0$ is a special point,
governed by a separate symmetry class.
Although our theory is limited to a quasi one-dimensional geometry, it
does show the importance of the chiral symmetry at the band center
$\varepsilon = 0$ and may thus contribute to the debate about the
localization properties of the random flux problem in higher spatial
dimensions.

This paper was motivated by two recent works.  First, in a recent paper,
one of the authors\cite{Furusaki98} computed $\langle g \rangle$ and $\mbox{var}\, g$
numerically for the random flux model in a wire geometry to a very high
accuracy.  While for nonzero energies $\varepsilon$, the result was
found to agree with analytical calculations for the unitary symmetry
class,\cite{Zirnbauer92,Mirlin94,Frahm95} for $\varepsilon=0$ a clear
difference with the unitary symmetry class was observed.  Second, for
the chiral symmetry classes, a scaling equation for the distribution of
the transmission eigenvalues in a quasi one-dimensional geometry was
derived and solved exactly in the chiral unitary case by Simons,
Altland, and two of the authors.\cite{Brouwer98} This scaling equation
is the chiral analogue of the so-called Dorokhov-Mello-Pereyra-Kumar
(DMPK) equation,\cite{Dorokhov,MPK,BeenakkerReview} which describes
the three standard symmetry classes and was solved exactly in the
unitary case by Beenakker and Rejaei.\cite{BeenakkerRejaei} However,
for the chiral unitary case, analytical results for the $L$ dependence
of $\langle g \rangle$ and $\mbox{var}\, g$ were lacking, so that a
comparison between the theory and the numerical results of
Ref.\ \onlinecite{Furusaki98} was not possible. In the present work
this gap is bridged.

In a wire geometry, the chiral unitary universality class undergoes a
striking even-odd effect first noticed by Miller and
Wang:\cite{Miller96,Chan96} 
The conductance $g$ decays exponentially with the
length $L$ if the number of channels $N$ is even, while critical
behavior is shown if $N$ is odd,
even in the limit of large $N$ that we consider here.
In the latter case, the average
conductance $\langle g \rangle$ decays algebraically, while the
conductance fluctuations are larger than the mean. We analyze how the
even-odd effect follows from the exact solution of the Fokker-Planck
equation of Ref.\ \onlinecite{Brouwer98} and compare with numerical
simulations of the random flux model.

We close the introduction by pointing out that the random flux problem
is also relevant to some strongly correlated electronic systems.  
In both the Quantum Hall Effect at half-filling
\cite{Kalmeyer92,Halperin93} and high $T_c$
superconductivity,\cite{Nagaosa90,Altshuler92} strong electronic
correlations can be implemented by auxiliary gauge fields.  In this
context, the random flux problem captures the contributions from the
static transverse gauge fields. 
Notice that the chiral symmetry is not required on physical grounds 
both for the Quantum Hall Effect at half-filling
and for high $T_c$ superconductivity.
Another area of applicability for our
results is the passive advection of a scalar field
\cite{advection,Miller96,Chalker97} and non-Hermitean quantum
mechanics.\cite{Sommers88,Hatano96,Efetov97,Mudry98,Brouwer98}
Finally, the striking sensitivity of the localization properties
in the random flux problem to the parity of the number $N$ of channels
is remarkably similar to that 
of the low energy sector of a single antiferromagnetic spin-$N/2$ chain
to the parity of $N$,\cite{Haldabe83} on the one hand, or to the sensitivity 
of the low energy sector of $N$ coupled antiferromagnetic spin-1/2 chains
to the parity of $N$,\cite{Dagotto96} on the other hand.

The paper is organized as follows. The random flux problem in a wire
geometry is defined in section \ref{sec:The random magnetic flux
model}.  The average and variance of the conductance are calculated
analytically in section \ref{sec:Moments of the conductance}.
Analytical predictions are compared to the numerical simulations in
section \ref{sec:Numerical Simulations}. 
We conclude in Sec.\ \ref{sec:Conclusions}.

\section{The random magnetic flux model}
\label{sec:The random magnetic flux model} 
\label{sec:2}

\begin{figure}
\begin{picture}(240,240)
\put(  0,230){$({\rm a})$}
\put(  0,220){\line(1, 0){210}}\put(215,218){$j=3$}
\put(  0,190){\line(1, 0){210}}\put(215,188){$j=2$}
\put(  0,160){\line(1, 0){210}}\put(215,158){$j=1$}
%
%
\multiput(15,220)(30, 0){7}{\line(0,-1){ 60}}
\put(30,140){$m=0$}\put(180,140){$m=5$}
%
%
\put( 28,202){$0$}
\put( 51,202){$\Theta_{0,2}$}
\put( 81,202){$\Theta_{1,2}$}
\put(111,202){$\Theta_{2,2}$}
\put(141,202){$\Theta_{3,2}$}
\put(171,202){$\Theta_{4,2}$}
%
%
\put( 28,172){$0$}
\put( 51,172){$\Theta_{0,1}$}
\put( 81,172){$\Theta_{1,1}$}
\put(111,172){$\Theta_{2,1}$}
\put(141,172){$\Theta_{3,1}$}
\put(171,172){$\Theta_{4,1}$}
%
%
\put(  0,100){${\rm (b)}$}
\thicklines
\put(  0, 90){\line(1,0){240}}
\put(  0, 50){\line(1,0){240}}
\put(  0, 77){$  c^L_{\nu ,+} \longrightarrow$}
\put(200, 77){$\  \longrightarrow c^R_{\nu',+}$}
\put(  0, 58){$  c^L_{\nu ,-} \longleftarrow $}
\put(200, 58){$\  \longleftarrow  c^R_{\nu',-}$}
\put( 50, 50){\line(0,1){40}}
\put(190, 50){\line(0,1){40}}
\thinlines
\put(120, 40){\vector(-1,0){70}}
\put(120, 40){\vector( 1,0){70}}
\put(117, 30){$L$}
\put(50,85){\line(1,-1){35}}
\put(50,80){\line(1,-1){30}}
\put(50,75){\line(1,-1){25}}
\put(50,70){\line(1,-1){20}}
\put(50,65){\line(1,-1){15}}
\put(50,60){\line(1,-1){10}}
\put(50,55){\line(1,-1){ 5}}
\multiput(50,90)(5,0){21}{\line(1,-1){40}}
\put(190,85){\line(-1,1){ 5}}
\put(190,80){\line(-1,1){10}}
\put(190,75){\line(-1,1){15}}
\put(190,70){\line(-1,1){20}}
\put(190,65){\line(-1,1){25}}
\put(190,60){\line(-1,1){30}}
\put(190,55){\line(-1,1){35}}
%
\end{picture}
\caption
{
(a) Lattice with $N=3$ threaded by random magnetic fluxes 
$\Theta_{m,j}$ in the disordered region $0<m<M$.
(b) Quantum wire with a disordered region of length $L=Ma$.
Incoming amplitudes are $c^L_{\nu,+}$ and $c^R_{\nu',-}$, 
whereas outgoing amplitudes
are $c^L_{\nu,-}$ and $c^R_{\nu',+}$, $\nu,\nu'=1,\ldots,N$,
in the left and right leads, respectively. 
In a quasi one-dimensional geometry, $M\gg N$.
}
\label{fig:plaquettes}
\end{figure}
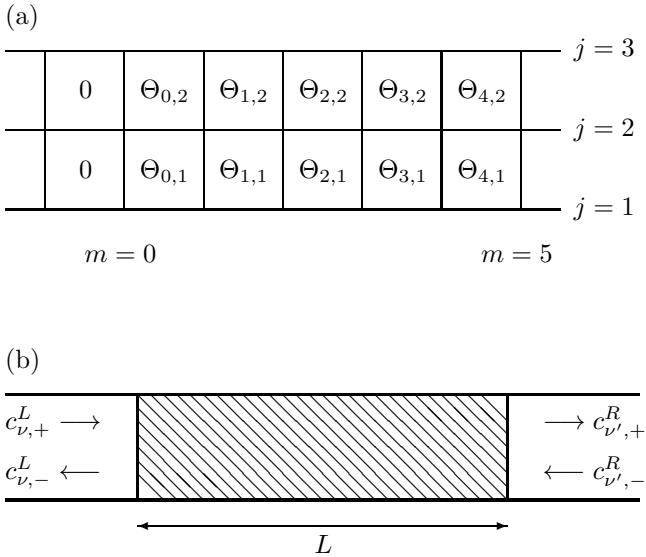

In the random flux model one considers a spinless electron on a
rectangular lattice in the presence of a random magnetic field with
vanishing mean. 
The magnetic field is perpendicular to the plane in which the electron
moves. 
In this paper, we study the random flux model in a wire
geometry and for weak disorder. 
This system is described by the Hamiltonian
\begin{eqnarray}
  {\cal H}\, \psi_{m,j} &=& -t [\psi_{m+1,j} + \psi_{m-1,j}] \nonumber \\
  && \mbox{} - t\, (1 - \delta_{j,N}) e^{i\theta_{m,j}} \psi_{m,j+1}
  \nonumber \\
  && \mbox{} - t\, (1 - \delta_{j,1}) e^{-i\theta_{m,j-1}} \psi_{m,j-1},
  \label{eq: def dynamics}
\end{eqnarray}
where $\psi_{m,j}$ is the wavefunction at the lattice site $(m,j)$,
labeled by the chain index $j=1,\ldots,N$ and by the column index $m$,
see Fig.\ \ref{fig:plaquettes}(a).  The Peierls phases $\theta_{m,j}$ result
from the flux $\Theta_{m,j} = \theta_{m+1,j} - \theta_{m,j}$ through
the plaquette between the sites $(m,j)$, $(m+1,j)$, $(m+1,j+1)$, and
$(m,j+1)$.  (The flux $\Theta_{m,j}$ does not uniquely determine all
the phases along all the bonds. We have used this freedom to choose the
nonzero phases along the transverse bonds only.)

We consider a system with Hamiltonian (\ref{eq: def dynamics}) where
the phases $\Theta_{m,j}$ take random values in a disordered strip 
$0 < m < M$ only, and are zero outside.\cite{Lee81}
We assume that the disordered region
is quasi one-dimensional, i.e., $M \gg N\gg1$, 
corresponding to a thick quantum wire. 
In the disordered region,
the Peierls phases $\theta_{m,j}$ are chosen at random in such a way
that the magnetic flux $\Theta_{m,j}=\theta_{m+1,j}-\theta_{m,j}$ is
uniformly distributed in $[-p\pi,p\pi]$ with $0<p\le1$. To be precise,
with $\theta_{m,j}$ given, $\theta_{m+1,j}$ is chosen from the interval
$[\theta_{m,j}-p\pi,\theta_{m,j}+p\pi]$ with uniform probability
$1/2p\pi$. The parameter $p$ controls the strength of disorder.
We assume weak disorder, i.e., $p\ll 1$.

The boundary conditions in the transverse directions that are implied by
the Hamiltonian (\ref{eq: def dynamics}) are ``open'', i.e., there are
no bonds between the chains $j=1$ and $j=N$. In this case, ${\cal H}$
has a special discrete symmetry, called the particle-hole or chiral
symmetry:  Under the transformation $\psi_{m,j}\to (-1)^{m+j}
\psi_{m,j}$, one has ${\cal H} \to - {\cal H}$. Hence, for each
realization of the random magnetic flux, the chiral symmetry ensures
that there exists an eigenstate of ${\cal H}$ with energy
$-\varepsilon$ for each eigenstate of ${\cal H}$ with energy
$+\varepsilon$.  Note that the band
center $\varepsilon=0$ is a special point.  The chiral symmetry is
broken by the addition of a random on-site potential to the Hamiltonian
(\ref{eq: def dynamics}). Another way to break the chiral symmetry is
to add bonds between the chains $j=1$ and $j=N$ and to impose periodic
boundary conditions in the transverse direction for $N$ odd.  
The presence of the chiral symmetry may have dramatic
consequences for charge transport through the disordered wire, as we
shall see in more detail in the next sections.

In order to find the conductance $g$ of the disordered region with the
random flux, we first compute the transfer matrix ${\cal M}$. To the
left and to the right of the disordered region, the wavefunction
$\psi_{m,j}$ that solves the Schr\"odinger equation ${\cal H} \psi =
\varepsilon \psi$ can be written as a sum of plane waves moving to the
right $(+)$ and to the left $(-)$,
\begin{eqnarray*}
  \psi_{j,m} &=&  
    \sum_{\nu=1}^{N_c} \sum_{\pm } c_{\nu,\pm}^{L} 
{e^{\pm i k_{\nu} m} \over \sin k_{\nu}}\, 
  \sin{ \nu j \pi \over N+1},\ \ m < 0 , \\
  \psi_{j,m} &=&  
    \sum_{\nu=1}^{N_c} \sum_{\pm } c_{\nu,\pm}^{R} 
{e^{\pm i k_{\nu} m} \over \sin k_{\nu}}\, 
  \sin{\nu j \pi \over N+1},\ \ m > M .
\end{eqnarray*}
where $\cos k_\nu = -\varepsilon/2t - \cos [\nu \pi/(N+1)]$. The
prefactor $1/\sin k_{\nu}$ is chosen such that an equal current is carried
in each channel. The number $N_c$ is the total number of propagating
channels at the energy $\varepsilon$, i.e., the total number of real
wavevectors $k_{\nu}$. We are interested in the transport
properties for $\varepsilon$ close to $0$, where $N_c = N$, and ignore
the distinction between $N_c$ and $N$ henceforth. The
coefficients $c_{\nu,\pm}^{L}$ and $c_{\nu,\pm}^{R}$ are related by the
transfer matrix ${\cal M}$ [see Fig.\ \ref{fig:plaquettes}(b)], 
\begin{equation}
  {{ c_{\nu,+}^{R}} \choose {c_{\nu,-}^{R} }} = \sum_{\nu'=1}^{N}
    {\cal M}_{\nu,\nu'} 
  {{ c_{\nu',+}^{L}} \choose {c_{\nu',-}^{L} }}. \label{eq:Mdef}
\end{equation}
Note that ${\cal M}_{\nu,\nu'}$ 
is a $2 \times 2$ matrix in Eq.\ (\ref{eq:Mdef}). Current conservation requires
\begin{equation}
  {\cal M} \Sigma_3 {\cal M}^{\dagger} 
  =
  \Sigma_3, \label{eq:CurCons}
\end{equation}
where $\Sigma_3 = \sigma_3 \otimes \openone_{N}$, $\sigma_3$ being the Pauli 
matrix and $\openone_N$ the $N \times N$ unit matrix.
In addition, at the special point $\varepsilon=0$, the chiral symmetry
of the Hamiltonian (\ref{eq: def dynamics}) results in the additional
symmetry
\begin{equation}
  \Sigma_1 {\cal M} \Sigma_1 = {\cal M}, \label{eq:ChiralSym}
\end{equation}
where $\Sigma_1 = \sigma_1 \otimes \openone_{N}$.

The eigenvalues of ${\cal M}{\cal M}^{\dagger}$, which occur in inverse
pairs $\exp(\pm 2 x_j)$, determine the transmission eigenvalues $T_j =
1/\cosh^2 x_j$ and hence the dimensionless
conductance $g$ through the Landauer formula\cite{Landauer70,Fisher81}
\begin{equation} \label{eq:Landauer}
  g = \sum_{j=1}^{N} T_j = \sum_{j=1}^{N} {1 \over \cosh^2 x_j}.
\label{eq: landauer formula}
\end{equation}
In the absence of disorder, all exponents $x_j$ are zero, 
and conduction is perfect, $g=N$. On the other hand, transmission 
is exponentially suppressed if all $x_j$'s are larger than unity. 
The smallest $x_j$ determines the localization properties of the
quantum wire.

For the quasi one-dimensional geometry $M\gg N\gg 1$ 
that we consider here and 
on length scales much larger than the mean free path associated to the
random magnetic field, the microscopic details of the microscopic
Hamiltonian ${\cal H}$ should no
longer be important.
Rather, the crucial ingredients are the symmetries
of ${\cal H}$. For nonzero energy, the only symmetry of ${\cal M}$ is given by
current conservation, Eq.\ (\ref{eq:CurCons}). In this case, for quasi
one-dimensional systems with sufficiently weak disorder, the
probability distribution $P(x_1,\ldots,x_N;L)$ of the parameters $x_j$ is
governed by the so-called Dorokhov-Mello-Pereyra-Kumar (DMPK)
equation,\cite{Dorokhov,MPK,BeenakkerReview}
\begin{mathletters} \label{eq:DMPK}
\begin{eqnarray} \label{eq:DMPK1}
  \ell {\partial P \over \partial L} &=&
  {1 \over 4 N} \sum_{j=1}^{N} {\partial \over \partial x_j}
  \left[ J {\partial \over \partial x_j} (J^{-1} P) \right], \\
  J &=& \prod_{k > j} |\sinh^2 x_j - \sinh^2 x_k|^2 \prod_{k} |\sinh(2 x_j)|.
  \label{eq:DMPK2}
\end{eqnarray}
\end{mathletters}%
Here $L = M a$ is the length of the disordered region, $a$ being the
lattice constant.
The mean free path $\ell$ depends on the disorder strength and on the
details of the microscopic model. 
The derivation of Eq.\ (\ref{eq:DMPK}) assumes $\ell\gg \lambda$,
$\lambda$ being the wave length at the Fermi energy.
The initial condition corresponding
to perfect transmission at $L=0$ is $P(x_1,\ldots,x_N;0) = \prod_{j}
\delta(x_j)$.  The Fokker-Planck equation (\ref{eq:DMPK}) describes the
unitary symmetry class.  For $\varepsilon=0$, in addition to current
conservation, the chiral symmetry (\ref{eq:ChiralSym}) has to be taken
into account. In Ref.\ \onlinecite{Brouwer98} it was shown that 
for weak disorder ($p\ll1$) the
distribution $P(x_1,\ldots,x_N;L)$ satisfies again a Fokker-Planck
equation of the form (\ref{eq:DMPK}), but with a different Jacobian
$J$,\cite{footnote}
\begin{mathletters} \label{eq:DMPKChiral}
\begin{eqnarray} \label{eq:DMPKChiral1}
  \ell {\partial P \over \partial L} &=&
  {1 \over 2 N} \sum_{j=1}^{N} {\partial \over \partial x_j}
  \left[ J {\partial \over \partial x_j} (J^{-1} P) \right], \\
  J &=& \prod_{k > j} |\sinh(x_j-x_k)|^2. \label{eq:Jchiral}
\end{eqnarray}
\end{mathletters}%
This Jacobian describes the chiral unitary symmetry class. As was shown
in Ref.\ \onlinecite{Brouwer98}, and as we shall see in more detail in the
next section, as a result of the replacement of the Jacobian
(\ref{eq:DMPK2}) by the Jacobian (\ref{eq:Jchiral}), 
the statistical distribution and the $L$-dependence of the conductance
$g$ at energy $\varepsilon = 0$ is quantitatively and qualitatively
different from that away from $\varepsilon = 0$.  In
Ref.\ \onlinecite{Brouwer98} it was shown that there exists a quantum
critical point induced by the randomness when $N$ is odd within the
chiral unitary symmetry class.
Away from zero energy, the transport properties of the disordered wire
are those expected from the standard unitary symmetry class.  A derivation of
Eq.\ (\ref{eq:DMPKChiral}) is given in Appendix \ref{ap: Derivation of
FP}.

The physical picture underlying Eqs.\ (\ref{eq:DMPK}) and
(\ref{eq:DMPKChiral}) is that the parameters $x_j$ undergo a ``Brownian
motion'' as the length $L$ of the disordered region is increased. The
Jacobian $J$ describes the ``interaction'' between the parameters $x_j$
in this Brownian motion process.  The key difference between the
unitary case and the chiral unitary case is the presence of an
interaction with ``mirror imaged'' eigenvalues $x_j$ in
Eq.\ (\ref{eq:DMPK2}), which is absent in Eq.\ (\ref{eq:Jchiral}).
To see this, we note that
both for the unitary and for the chiral unitary cases, the Jacobian $J$
vanishes if a parameter $x_j$ coincides with $x_k$, $k \neq j$.
However, in the unitary case (\ref{eq:DMPK2}), $J$ also vanishes if
$x_j$ coincides with a mirror image $-x_k$, $k \neq j$, or if $x_j = 0$
(i.e., $x_j$ coincides with its own mirror image).  The vanishing of the
Jacobian $J$ implies a repulsion of the parameters $x_j$
in the underlying Brownian motion process.  Hence, whereas $x_j$ feels
a repulsion from the other $N-1$ parameters $x_k$, $k\neq j$, in the
chiral unitary case (\ref{eq:DMPKChiral}), $x_j$ feels an additional
repulsion from the $N-1$ mirror images $-x_k$, $k\neq
j$, and from its own mirror image $-x_j$ in the standard unitary case
(\ref{eq:DMPK}).

It can be shown\cite{BeenakkerReview,Brouwer98} that the parameters
$x_j$ repel each other by a constant force in the large-$L$ limit,
irrespective of their separation.  This long-range repulsion results in
the so-called ``crystalization of transmission eigenvalues'': The
fluctuations of the parameters $x_j$ are much smaller than the spacings
between their average positions.\cite{BeenakkerReview}
Away from zero energy, i.e., in the unitary symmetry class, all $x_j$
can be chosen positive because of repulsion from their mirror images,
and their average positions are\cite{BeenakkerReview}
\begin{equation}
  \langle x_j \rangle = (2 j -1) L/2 N \ell,\ \ j=1,\ldots,N.
\end{equation}
In the chiral unitary symmetry class, the $x_j$ can be both positive
and negative since there is no repulsion from the mirror images, and
one has\cite{Brouwer98}
\begin{equation}
  \langle x_j \rangle = (N+1-2j)L/ N \ell,\ \ j=1,\ldots,N.
\end{equation}
In the unitary symmetry class and in the chiral unitary class with even
$N$ the net force on each parameter $x_j$ is finite, and they grow
linearly with the length $L$. Hence, by Eq.\ (\ref{eq:Landauer}), the
conductance $g$ is exponentially suppressed for $L \gg N \ell$.
However, for the chiral disordered wire with an odd number of channels
$N$, the net force on the middle eigenvalue $x_{(N+1)/2}$ vanishes: it
remains in the vicinity of the origin and the conductance is not
exponentially suppressed.\cite{Brouwer98}
Thus, the quantum wire with random flux with an odd number $N$ of channels
goes through a quantum critical point at zero energy whereas it remains
non-critical for an even number $N$ of channels. A more quantitative
description of this even-odd effect is developed in the next section.

\section{Moments of the conductance} \label{sec:3}
\label{sec:Moments of the conductance}

\subsection{Method of bi-orthonormal functions}
\label{subsec: bi-ortho}

To calculate the moments of the conductance $g$, we
make use of the exact solution of the Fokker-Planck equation
(\ref{eq:DMPKChiral}),\cite{Brouwer98} 
\begin{eqnarray}
P(x_1,\ldots,x_N;L) &\propto& 
\prod_{j=1}^Ne^{-{N{\ell}\over 2 L}x^2_j}
\nonumber\\
&& \mbox{} \times
\prod_{j<k}
(x_j-x_k)\sinh(x_j-x_k).
\label{eq: jpd if unitary}
\end{eqnarray}
The proportionality constant is fixed by normalization of the probability distribution. A derivation of Eq.\ (\ref{eq: jpd if unitary}) is
presented in Appendix \ref{ap: P({x_j})}.

The moments of $g$ can be computed from the $n$-point correlation
functions\cite{Mehta} 
\begin{eqnarray}
&&
R_{n}(x_1,\ldots,x_n;L)=
\\
&&
{N!\over(N-n)!}
\int_{-\infty}^{+\infty} dx_{n+1}\ldots
\int_{-\infty}^{+\infty} dx_{  N}
P(x_1,\ldots,x_N;L),
\nonumber
\end{eqnarray}
and the Landauer formula (\ref{eq:Landauer}). For example, the first and
second moments of $g$ are
\begin{mathletters}
\begin{eqnarray}
\langle g\rangle &=&
\int_{-\infty}^{+\infty}dx\,{R_{1}(x;L)\over\cosh^2x},
\\
\langle g^2\rangle &=&
\int_{-\infty}^{+\infty}\!\!dx_1 
\int_{-\infty}^{+\infty}\!\!dx_2
{R_{2}(x_1,x_2;L)\over\cosh^2x_1\cosh^2x_2} \nonumber
\\
  && \mbox{} +
\int_{-\infty}^{+\infty}dx\,{R_{1}(x;L)\over\cosh^4x}.
\end{eqnarray}
\end{mathletters}%
Here we compute $R_{n}(x_1,\ldots,x_n;L)$ using the method of
bi-orthonormal functions developed by 
Muttalib\cite{Muttalib95} and Frahm\cite{Frahm95}
for a disordered wire in the unitary symmetry class.
The idea is to construct, for any
given $N$ and $L$, a function $K_{L}(x,y)$ with the following
properties,
\begin{mathletters}
\label{eq:hypo}
\begin{eqnarray}
&&
\int_{-\infty}^{+\infty} dx\, K_{L}(x,x)=N, 
\label{eq: assumption i}\\
&&
\int_{-\infty}^{+\infty} dy\, K_{L}(x,y)K_{L}(y,z)=K_{L}(x,z),
\label{eq: assumption ii} 
\\
&&
P(\{x_i\};L)= c_N\, \det\, 
\left[ K_{L}(x_i,x_j) \right]_{i,j=1,\ldots,N}.
\label{eq: assumption iii}
\end{eqnarray}
\end{mathletters}
If such a function exists, 
it is known from random matrix theory\cite{Mehta} 
that $c_N = 1/N!$ and
\begin{eqnarray}
\label{eq:thm}
 R_{n}(\{x_i\};L) &=&{\rm det}\, 
\left[ K_{L}(x_i,x_j) \right]_{i,j=1,\ldots,n}.
\end{eqnarray}

Our construction of the function $K_{L}(x,y)$ starts with a
representation of $P(x_1,\ldots,x_N;L)$ in Eq.\ (\ref{eq: jpd if
unitary}) as a product of two determinants. Making use of the
identities
\begin{eqnarray*}
  \prod_{j<k} (x_k - x_j) &=& \det\left[x_k^{j-1}\right]_{j,k=1,\ldots,N}, \\
  \prod_{j<k} \sinh(x_k - x_j) &=&
    \det\left[\case{1}{2}e^{(N+1-2j)x_k}\right]_{j,k=1,\ldots,N},
\end{eqnarray*}
we find
\begin{mathletters} \label{eq:P as product 2 Vand}
\begin{eqnarray}
P(\{x_i\};L) & \propto &
{\rm det}\, \left[\phi_j   (x_k)\right]_{j,k=1,\ldots,N}
\nonumber \\&& \mbox{} \times
{\rm det}\, \left[\eta_j(x_k)\right]_{j,k=1,\ldots,N},
\end{eqnarray}
where
\begin{eqnarray}
&&
\phi_j(x)= x^{j-1},
\\
&&
\eta_j(x)=e^{-{N{\ell}\over 2 L} x^2 +(N+1-2j)x}.
\end{eqnarray}
\end{mathletters}%
Note that the way we write $P$ as a product of two determinants in
Eq.\ (\ref{eq:P as product 2 Vand}) is not unique. In particular, we
are free to replace the sets of functions $\{\phi_j\}$ and $\{\eta_j\}$
by an arbitrary set of linear combinations $\{\tilde \phi_j\}$ and
$\{\tilde \eta_j\}$. This freedom is crucial for the construction of
the function $K_{L}(x,y)$, as we shall see below.

Since the product of two determinants equals the determinant of the
product of the corresponding matrices and since transposition of a
matrix leaves the determinant unchanged, it is tempting to identify
$K_{L}(x,y)$ with $\sum_{j=1}^N\phi_j(x)\eta_j(y)$. In this way, Eq.\
(\ref{eq: assumption iii}) is satisfied. However, with
this choice, the remaining two conditions (\ref{eq: assumption i}) and
(\ref{eq: assumption ii}) are not obeyed. This problem can be
solved by making use of the above-mentioned freedom to
replace the sets of functions $\{\phi_j\}$ and $\{\eta_j\}$ by linear
combinations $\{\tilde \phi_j\}$ and $\{\tilde \eta_j\}$. One easily
verifies that if we choose these linear combinations such that they
are bi-orthonormal,\cite{Muttalib95}
\begin{eqnarray}
\int_{-\infty}^{+\infty} dx\, 
\tilde\phi_{j}(x)\tilde\eta_{k}(x)=\delta_{jk},
\qquad j,k=1,\ldots,N, \label{eq: bi-ortho}
\end{eqnarray}
all three conditions (\ref{eq:hypo}) are met if we set
\begin{equation}
  K_{L}(x,y) = \sum_{j=1}^N \tilde \phi_j(x) \tilde \eta_j(y).
\end{equation}
The construction of the bi-orthonormal functions 
$\tilde\phi_j$ and $\tilde \eta_j$ is done below.

First, we define the set $\{\tilde\eta_{j}(x)\}$, 
$j=1,\ldots,N$,
by completing the square in the exponent of $\eta_j(x)$
and then normalizing $\eta_j(x)$,
\begin{equation}
\tilde\eta_{j}(x)=\sqrt{1 \over 2 \pi \sigma}\, 
e^{-\left(x-{ \varepsilon_{j} \sigma }\right)^2/2 \sigma},
\end{equation}
where we abbreviated
\begin{eqnarray}
  \sigma = L/N \ell,\qquad
  \varepsilon_j &=& N + 1 - 2 j.
\end{eqnarray}

The functions $\tilde \phi_j$, being linear combinations of $\phi_j(x)
= x^{j-1}$, are polynomials themselves, too. Their (maximal) degree is
$N-1$. As a first step
towards their construction, we define the
polynomials
\begin{equation} \label{eq:qdef}
  p_j(x) = \sqrt{1\over 2 \pi \sigma}
  \int_{-\infty}^{\infty} dy\, (i y/\sigma)^{j-1} e^{-(y + i x)^2/2\sigma},
\end{equation}
which satisfy the special property
\begin{equation} \label{eq:special}
  \int_{-\infty}^{\infty} dx\, p_j(x) \tilde \eta_k(x) = (\varepsilon_k)^{j-1}.
\end{equation}
Notice that $p_j(x)$ is of degree $j-1$.  According to
Eq.\ (\ref{eq:special}), the overlap matrix between the polynomials
$p_j$ and the Gaussians $\tilde\eta_j$ is independent of $L$.
Construction of bi-orthonormal functions $\tilde\phi_j$ and
$\tilde\eta_j$ is thus achieved by choosing $L$-independent linear
combinations of the polynomials $p_j$ that diagonalize the overlap
matrix (\ref{eq:special}).
This is done using the Lagrange interpolation polynomials\cite{Frahm95}
\begin{equation}
L_{m}(x)=
\prod_{n\neq m}
{x-\varepsilon_{n}\over\varepsilon_{m}-\varepsilon_{n}},
\label{eq: Lagrange polynomial}
\end{equation}
which are of degree $N-1$ and obey 
$L_{m}(\varepsilon_{n}) = \delta_{m,n}$. 
We infer that the desired polynomials $\tilde \phi_j(x)$ are given by
\begin{equation}
\tilde\phi_{j}(x)= \sqrt{1 \over 2 \pi \sigma}
\int_{-\infty}^{+\infty}dy\, 
L_{j}({i} y/\sigma)\, 
e^{- \left(y+ {i}x\right)^2/2 \sigma}. \label{eq:phitildedef}
\end{equation}

Putting everything together, we find that
\begin{eqnarray} \label{eq: main result of model I}
K_{L}(x,z) &=&
{1 \over 2 \pi \sigma} \sum_{j=1}^{N}
\int_{-\infty}^{+\infty}dy\,
L_{j}({i}y/\sigma)\,
\nonumber \\
&& \mbox{} \times\,
\exp \left[ - {\left(y+{i}x\right)^2 + 
               \left(z- {\varepsilon_{j}} \sigma \right)^2 \over 2 \sigma}
     \right].
\label{eq: K_{L}}
\end{eqnarray}
Now, moments of the conductance $g$ can be calculated 
with the help of Eq.\ (\ref{eq:thm}). In particular, we find that the
average and variance of $g$ are given by
\begin{eqnarray}
\langle g\rangle&=&
\int_{-\infty}^{+\infty}\!\! dx\,{K_{L}(x,x)\over\cosh^2 x}, 
\label{eq: g}
\\
\mbox{var}\, g &=& -
\int_{-\infty}^{+\infty}\!\! dx_1
\int_{-\infty}^{+\infty}\!\! dx_2\,
{K_{L}(x_2,x_1)K_{L}(x_1,x_2)\over\cosh^2 x_1 \cosh^2 x_2 } \nonumber \\
  && \mbox{} + 
\int_{-\infty}^{+\infty}\!\! dx\,{K_{L}(x,x)\over\cosh^4 x}. \label{eq: varg}
\end{eqnarray}

\subsection{Average conductance}

After some shifts of integration variables and with the help of the
Fourier transform of $\cosh^{-2}x$,
\begin{eqnarray}
\int_{-\infty}^{+\infty}dx\,{e^{{i}yx}\over\cosh^2x}&&=
{\pi y\over\sinh(\pi y/2)}
\nonumber\\
&&=
2\prod_{k=1}^\infty\,\left(1+{y^2\over4k^2}\right)^{-1},
\label{eq: infinite product x^(-1)sinh x}
\end{eqnarray}
we obtain from Eqs.\ (\ref{eq: main result of model I}) and (\ref{eq: g})
an expression for the average conductance $\langle g \rangle$ 
at the energy $\varepsilon=0$ that involves one integration
and one (finite) summation only,
\begin{eqnarray}\label{eq: mean g finite N}
\langle g \rangle & = &
\sum_{m=1}^{N}\, c_{m}\, e^{- \varepsilon_m^2 \sigma /2 },
\label{eq:finite series for mean g} \\
c_{m} &=&
  \int_{-\infty}^{+\infty}\!\!\!\! dy\,
{
L_{m}(\varepsilon_{m}-{i}y)\,
y\, e^{- (y+{i}\varepsilon_{m})^2 \sigma /2}\,
\over2\sinh(\pi y/2) \nonumber
},
\end{eqnarray}
where, as before, $\sigma = L/N \ell$.
In the limit $N \gg 1$ at fixed $\sigma$ (the so-called thick-wire
limit), Eq.\ (\ref{eq: mean g finite N}) can be further simplified.
Hereto we use the second identity of Eq.\ (\ref{eq: infinite product
x^(-1)sinh x}) to cancel the Lagrange interpolation
polynomial in the coefficient $c_m$,
\begin{eqnarray}
  c_{m} &=&
\int_{-\infty}^{+\infty} {dy\over\pi} e^{-{y^2 \sigma/2}}
\prod_{k\in\Lambda_{m}}    
\left(1-{{i}y+\varepsilon_{m}\over2k}\right)^{-1},
\nonumber\\
\Lambda_{m} &=&
{\bf Z}-\{-m+1,\ldots,-m+N\}. \label{eq:cminf}
\end{eqnarray}
In the limit $N \to \infty$, only $m$'s close to $(N+1)/2$ contribute
to $\langle g \rangle$.  For those $m$, we may replace the infinite
product on the r.h.s.\ of Eq.\ (\ref{eq:cminf}) by unity, and find $c_m
= (2/\pi \sigma)^{1/2}$. 
Hence, for $N \gg 1$ even,
\begin{mathletters}
\label{eq: mean g in large N limit}
\begin{equation}
\langle g\rangle=
\sqrt{2\over \pi \sigma}\, \vartheta_2(0|2 {i} \sigma/\pi)\equiv
\sqrt{2\over \pi \sigma}\sum\limits_{{m\in{\bf Z}\atop m\, {\rm odd}}}
e^{-{m^2 \sigma/2}},
\label{eq: mean g in large N even limit}
\end{equation}
whereas for $N \gg 1$ odd,
\begin{equation}
\langle g\rangle=
\sqrt{2\over \pi \sigma}\, \vartheta_3(0|2 {i} \sigma/\pi)\equiv
\sqrt{2\over \pi \sigma}\sum\limits_{{m\in{\bf Z}\atop m\, {\rm even}}}
e^{-{m^2 \sigma/2}}.
\end{equation}
\end{mathletters}%
Here $\vartheta_2$ and $\vartheta_3$ are
the Jacobi's theta functions.\cite{JacobiTheta}

The dramatic difference between even and odd channel numbers discovered
in Refs.\ \onlinecite{Miller96,Brouwer98} follows immediately
from Eqs.\ (\ref{eq:finite series for mean g}) or
(\ref{eq: mean g in large N limit}) in the regime $L \gg N
\ell$. For even $N$, each term in the summation
decays exponentially with $L$. The exponential decay of $\langle g
\rangle$ is governed by the slowest decaying terms in the summation in
Eq.\ (\ref{eq:finite series for mean g}), i.e., the contributions from
$\varepsilon_m = \pm 1$, i.e., from
$m = N/2$ or $m=N/2+1$. Hence for $L \gg N
\ell$ we find
\begin{equation}
  \langle g \rangle \approx 
\sqrt{8 \xi \over \pi L}\, e^{-L/2 \xi},\qquad
  \xi = N\ell,\qquad \mbox{$N$ even}. 
\label{eq:Geven}
\end{equation}
Eq.\ (\ref{eq:Geven}) allows us to identify $\xi$ as the localization
length.\cite{BeenakkerReview}
For odd $N$, there is one term in the summation 
(\ref{eq:finite series for mean g}) 
that does not decay exponentially with $L$. It is the
contribution from the channel with $\varepsilon_m = 0$, $m = (N+1)/2$.
In this case, we again define $\xi=N\ell$,
although it is now merely a crossover length scale,
to be the characteristic length scale above
which the slow algebraic decay of $\langle g\rangle$ sets in,
\begin{equation}
  \langle g \rangle \approx \sqrt{2 \xi \over \pi L},
\qquad \mbox{$N$ odd},\qquad L\gg\xi.
\label{eq:Godd}
\end{equation}
In Fig.\ \ref{fig:<g> I,finite N} we have shown the average conductance
for $N=1,2,3,4$ as a function of $L/\xi$ and the asymptotic result
for large $N$.

\narrowtext
\begin{figure}
\centerline{\epsfxsize=3in\epsfbox{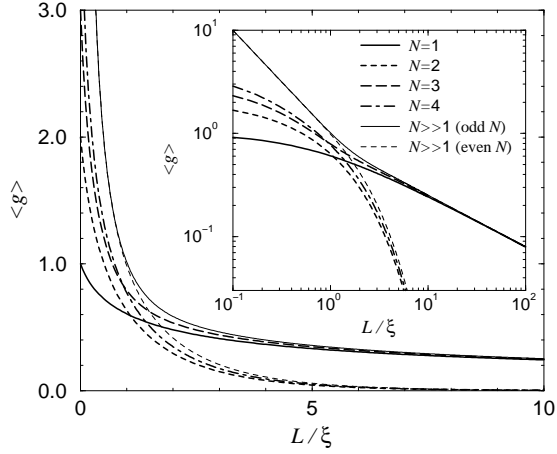}}
\caption{\label{fig:<g> I,finite N}
Average conductance at zero energy 
as a function of $L/\xi$ ($\xi=N\ell$)
for a quantum wire with a finite number of channels $N=1,2,3,4$
and in the chiral unitary symmetry class.
The even-odd effect is clearly visible for $L/\xi\gg1$:
(i)  Exponential decay of $\langle g\rangle$ for $N$ even,
(ii) Algebraic   decay of $\langle g\rangle$ for $N$ odd.
}
\end{figure}

To study the average conductance in the diffusive regime 
${\ell}\ll L\ll \xi$, we use the Poisson summation formula
\begin{mathletters}
\begin{eqnarray}
&&
\sum_{m\in{\bf Z}}\delta(x-2m-1)=
\frac{1}{2}\sum_{n\in{\bf Z}}e^{{i}\pi n(x-1)},
\label{eq:Poisson even N}
\\
&&
\sum_{m\in{\bf Z}}\delta(x-m)=
\sum_{n\in{\bf Z}}e^{2\pi{i}nx},
\label{eq:Poisson odd  N}
\end{eqnarray}
\end{mathletters}
to convert Eq.\ (\ref{eq: mean g in large N limit})
into 
\begin{eqnarray}
\label{eq: mean g in diffusive regime, no appro}
\langle g\rangle &=& {\xi \over L} 
\\
  && \mbox{} +
\cases{
{2 \xi \over L} \sum\limits_{n=1}^{\infty} 
(-1)^ne^{-\pi^2n^2 \xi/2L},& 
$N \gg 1$ even,\cr
&\cr
{2 \xi \over L} \sum\limits_{n=1}^{\infty} 
e^{-\pi^2n^2 \xi/2L},& 
$N \gg 1$ odd.
} \nonumber
\end{eqnarray}
Hence the even-odd effect is non-perturbative in $L/\xi$ and we
see that $\xi = N \ell$ is the characteristic
length scale at which the even-odd effect shows up.
Whereas the leading terms are identical, the first non-perturbative
correction to $\langle g\rangle$ differs by a sign for even and odd $N$.

In Fig.\ \ref{fig:<g> I,infinite N} we plot the average conductance for $N
\gg 1$ as a function of $L/\xi$ and compare with the unitary symmetry
class, which is appropriate for energies $\varepsilon \neq 0$. In the
unitary symmetry class, $\langle g \rangle$ decays
exponentially,\cite{Zirnbauer92,Mirlin94,Frahm95} irrespective of the
parity of $N$, but with a different localization length $\xi_{\rm u}$,
\begin{equation}
  \langle g \rangle \propto e^{-L/2 \xi_{\rm u}},\ \ \xi_{\rm u} = 2 N \ell.
\end{equation}
The unitary symmetry class is appropriate for the random flux model if
the energy $\varepsilon$ is nonzero. Hence, moving the energy
$\varepsilon$ away from zero causes a factor $2$ increase in the
localization length if the number of channels is even, and a dramatic
decrease in the average conductance if $N$ is odd.

\begin{figure}
\centerline{\epsfxsize=3in\epsfbox{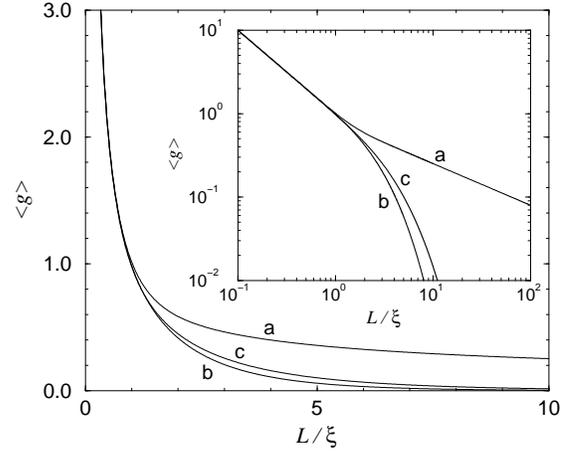}}
\caption{\label{fig:<g> I,infinite N}
Average conductance as a function of $L/\xi$ for a quantum wire in the
limit $N\gg1$.  Curves (a) and (b) are for large odd and even $N$ at
zero energy, when the system is in the chiral unitary symmetry class.
Curve (c) shows $\langle g \rangle$ for a quantum wire in the unitary
symmetry class with the same mean free path as in (a) and (b).  
}
\end{figure}

\subsection{Variance of the conductance}

\begin{figure}
\centerline{\epsfxsize=3in\epsfbox{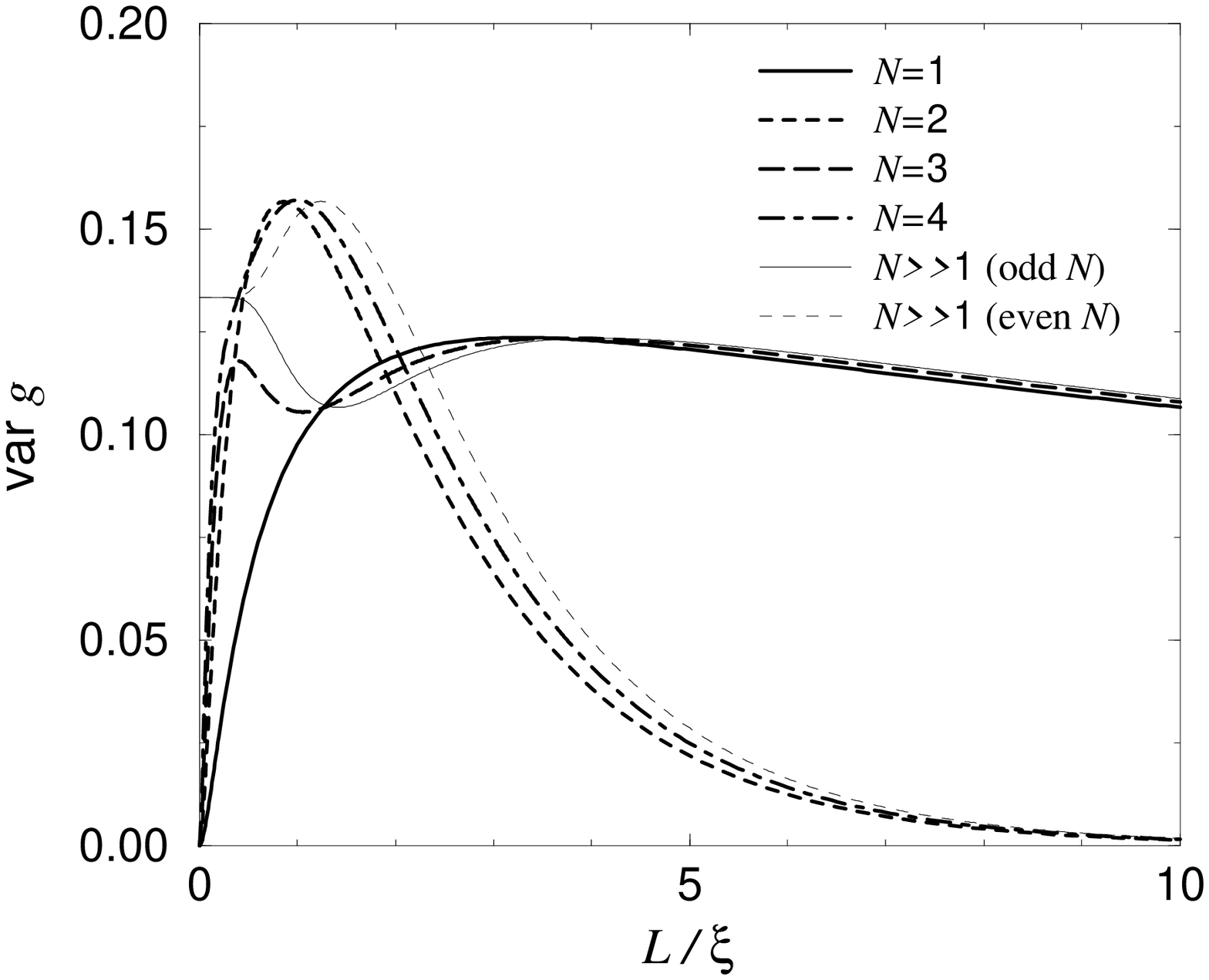}}
\smallskip
\caption{\label{fig: var g for finite N} 
Same as Fig.\ \protect\ref{fig:<g> I,finite N}, but for the variance
of the conductance.
}
\end{figure}

Proceeding as in the previous subsection, we find from Eqs.\ (\ref{eq:
main result of model I}), (\ref{eq: varg}), and (\ref{eq: infinite
product x^(-1)sinh x}),
\begin{eqnarray} \label{eq: var g finite N}
{\rm var}\, g &=&
\sum_{m,n=1}^{N}
c_{m,n}\, e^{-(\varepsilon_{m}^2 + \varepsilon_{n}^2) \sigma/2}
 \nonumber \\ && \mbox{}
+ \sum_{m=1}^{N}\, c'_{m}\, e^{-{\varepsilon_{m}^2 \sigma/2}},
\end{eqnarray}
with the coefficients
\begin{eqnarray*}
c_{m,n} &=& -
\int_{-\infty}^{\infty} dy_1\,
{y_1\, L_{m}(\varepsilon_{n}-{i}y_1)\, 
e^{-{(y_1+{i}\varepsilon_{n})^2 \sigma/2}}
 \over2\sinh\left(\pi y_1 / 2\right)}
\nonumber\\
&& \mbox{} \times
\int_{-\infty}^{\infty} dy_2\, 
{y_2\, L_{n}(\varepsilon_{m}-{i}y_2)\,
e^{-{(y_2+{i}\varepsilon_{m})^2 \sigma/2}}
 \over 2\sinh\left(\pi y_2 / 2\right)}, \\
  c'_{m} &=&
\int_{-\infty}^{\infty} dy\, 
{(y^3+4y)\, L_{m}(\varepsilon_{m}-{i}y)\,
e^{-(y+{i}\varepsilon_{m})^2 \sigma/2}\,
\over12\sinh(\pi y/2)}. \nonumber
\end{eqnarray*}
We plot ${\rm var}\, g$, which is computed from
Eq.\ (\ref{eq: var g finite N}) for $N=1,2,3,4$, 
in Fig. \ref{fig: var g for finite N},
together with the thick wire limit $N\gg 1$.
The even-odd effect is clearly seen when $L/\xi\gg1$.

\narrowtext

In the limit $N \to \infty$ at fixed $N \ell/L$ further simplifications
are possible. We find
\begin{eqnarray}
  \mbox{var}\, g &=& 
-{\sum_{m=-\infty}^{\infty}}\!\!\!\!\! \left. \vphantom {M^M_M} \right.'\,
 {\sum_{n=-\infty}^{\infty}}\!\!\!\! \left. \vphantom {M^M_M} \right.'\,
 f_{m,n} f_{n,m} +
   {\sum_{m=-\infty}^{\infty}}\!\!\!\!\! \left. \vphantom {M^M_M} \right.'\,
 f'_{m}, \nonumber \\ 
f_{m,n} &=&
  \sqrt{2 \over \pi \sigma} e^{-m^2 \sigma/2} \nonumber \\ && \mbox{}
  + {1 \over 2} \left[(m-n)\, \mbox{erf}\, \left( m \sqrt{\sigma/2} \right)
  - | m - n | \right], \nonumber \\
  f'_{m} &=& \sqrt{1 \over 18 \pi \sigma} \left(4 - m^2 + \sigma^{-1} \right)
  e^{-m^2 \sigma/2}, \label{eq: var g, large N}
\end{eqnarray}
where the primed summations are restricted to even (odd) 
$m$ and $n$ for $N$ odd (even).
The error function ${\rm erf}(x)$ is defined as
\begin{eqnarray*}
{\rm erf}\, (x)= {2\over\sqrt{\pi}}\int_0^{x}dt\, e^{-t^2}.
\end{eqnarray*}

For $L\gg\xi$ Eq.\ (\ref{eq: var g, large N}) simplifies to
\begin{equation}
{\rm var}\ g \approx
\cases{
  \sqrt{\frac{2 \xi}{\pi L}}\, e^{-L/2 \xi},& $N \gg 1$ even,\cr
&\cr
  \sqrt{\frac{8 \xi}{9\pi L}},& $N \gg 1$ odd.\cr
}
\label{eq: var g in loc}
\end{equation}
The variance of the conductance decays exponentially for large even $N$
with the same decay length as the average $\langle g \rangle$, while
$\mbox{var}\, g$ decays algebraically for large odd $N$.
Note that $\langle g \rangle$ and $\mbox{var}\, g$ decay with the same
power of $L$.

After some tedious algebra starting from Eq.\ (\ref{eq: var g, large N})
to extract an expression well suited for an asymptotic expansion in
small $L$, we find for the diffusive regime ${\ell}\ll L\ll\xi$,
\begin{eqnarray}
{\rm var}\, g &=& \nonumber \cases{ {2 \over 15} +
  \frac{\pi^2}{3} \left({\xi \over L} \right)^3 
e^{-{\pi^2 \xi \over 2 L }}+\ldots,& $N \gg 1$ even
,\cr
&\cr {2 \over 15} -
  \frac{\pi^2}{3} \left({\xi \over L} \right)^3 
e^{-{\pi^2 \xi \over 2 L}}+\ldots,& $N \gg 1$ odd.
} \\ \label{eq: var g in diffusive regime}
\end{eqnarray}
Again, we see that the difference between even and odd channel numbers
shows up in terms that are non-perturbative in $L/\xi$.
The leading term $2/15$ in ${\rm var}\, g$ is universal and twice the value
of its counterpart for a disordered quantum wire in the unitary ensemble.\cite{Zirnbauer92,Mirlin94,Frahm95} Hence, moving the energy $\varepsilon$
away from zero decreases the conductance fluctuations by a factor two in
the diffusive regime. 
The factor two decrease of ${\rm var}\, g$ upon
breaking the chiral symmetry is reminiscent of the factor two difference 
for the conductance fluctuations between the standard orthogonal and unitary 
symmetry classes.\cite{BeenakkerReview}
The enhancement of the conductance fluctuations at $\varepsilon=0$  
had been observed previously in numerical
simulations of the two-dimensional random flux problem
by Ohtsuki et al.\cite{Ohtsuki93}

\begin{figure}
\centerline{\epsfxsize=3in\epsfbox{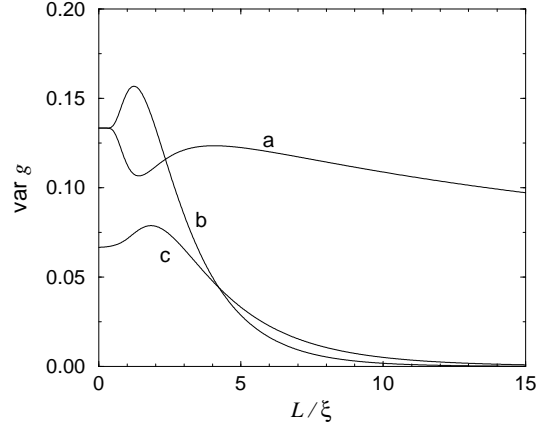}}
\smallskip
\caption{\label{fig: var g for large N}
Same as Fig.\ \protect\ref{fig:<g> I,infinite N}, but for the variance
of the conductance.}
\end{figure}

Figure \ref{fig: var g for large N} contains a plot of ${\rm var}\, g$
versus $L/\xi$, and offers a comparison with the unitary symmetry
class.\cite{Zirnbauer92,Mirlin94,Frahm95} In the unitary symmetry
class, $\mbox{var}\, g$ takes the universal value $1/15$ in the
diffusive regime $L \ll N \ell$, while $\mbox{var}\, g \propto \exp(-L/2
\xi_{\rm u})$, $\xi_{\rm u} = 2 N \ell$, in the localized regime $L \gg
N \ell$. 

\narrowtext

\section{Numerical simulations}
\label{sec:Numerical Simulations}

In this section we present numerical simulations for the conductance
$g$ in the random flux model (\ref{eq: def dynamics}). The average and
variance of $g$ were studied previously by Avishai {\em et
al.}\cite{Avishai93} and by Ohtsuki {\em et al.}\cite{Ohtsuki93}
for the random flux model in a square geometry. However, for a
comparison with the theory of Sec.\ \ref{sec:3} and to identify
the symmetry class it is necessary to study a wire geometry and
large system sizes. This is done below.

For each disorder configuration, we calculate the conductance using the
Landauer formula (\ref{eq: landauer formula}), which we use in the more
conventional form
\begin{equation}
g = \sum^{N_c}_{\mu, \nu=1} |{\bf t}_{\mu, \nu}|^2,
\label{eq: Landauer}
\end{equation}
Here $N_c$ is the number of propagating channels in the leads and ${\bf t}$
is the $N_c \times N_c$ transmission matrix, which relates the
amplitudes of the incoming and outgoing waves on the left and the right
of the disordered sample.  The eigenvalues $T_j$ of the matrix 
${\bf t}{\bf t}^{\dag}$ are the same as in Eq.\ (\ref{eq: landauer formula}).
(The simulations are aimed at energies $\varepsilon$ close to zero,
where $N_c$ equals $N$. Hence, as before, we drop the notational distinction
between $N_c$ and $N$).

The transmission matrix ${\bf t}$ is computed through the recursive Green
function method.\cite{MacKinnon83,Ando89,Baranger} In this method, $N \times
N$ matrix Green functions $F_{jk}$ for reflection and $G_{jk}$ for
transmission through the disordered region are computed using the
recursive rule,
\begin{eqnarray}
F(m+1)&=&[\varepsilon-H^m-t^2\,F(m)]^{-1},
\nonumber\\
G(m+1)&=&-t\,G(m)\,F(m+1), 
\end{eqnarray}
where the matrix elements of $H^m$ are 
\begin{eqnarray*}
H^m_{j,k} &=& \nonumber
-t(1-\delta_{j,N})\,e^{ i\theta_{m,j}}\delta_{j+1,k}
\\
&&
-t(1-\delta_{j,1})\,e^{-i\theta_{m,j-1}}\delta_{j-1,k}. 
\end{eqnarray*}
The initial conditions at $m=0$ are those of a Green function at the edge of
an isolated perfect lead:
\begin{eqnarray}
F_{jk}(0) &=& G_{jk}(0) \nonumber \\
  &=& -\frac{2}{N+1}\sum^N_{\nu=1}e^{i k_{\nu}} 
\sin {\nu j \pi \over N+1} \sin {\nu k \pi \over N+1},
\end{eqnarray}
where $\cos k_{\nu} = -\varepsilon/2t - \cos [\nu \pi/(N+1)]$, see
Sec.\ \ref{sec:2}. The scattering channels are those modes with real
wavevectors $k_{\nu}$.
The Green function that we need is obtained by taking into account the
perfect lead boundary condition on the right of the disordered region,
\begin{eqnarray*}
F(M)&=&\left[F(0)^{-1}-t^2\,F(M-1)\right]^{-1},\\
G(M)&=&-t\,G(M-1)\,F(M).
\end{eqnarray*}
The matrix Green function $G(M)$ describes the propagation from
$m=0$ to $m=M$. 
The absolute value of the transmission matrix element 
at energy $\varepsilon$ is then given by
\begin{eqnarray*}
|{\bf t}_{\mu,\nu}|^2 &=&
4\sin k_{\mu} \sin k_{\nu}
\nonumber\\
&&\times
\left|\frac{2}{N+1}\sum^N_{j,k=1} G_{jk}(M)
 \sin {\mu j \pi \over N+1} \sin {\nu k \pi \over N+1} \right|^2.
\end{eqnarray*}
This procedure is repeated for each disorder configuration, and the
average and variance of the conductance are obtained by taking an average
over $2\times 10^4$ samples. The transverse boundary conditions are
those of Eq.\ (\ref{eq: def dynamics}), i.e., open boundaries, unless
explicitly indicated otherwise. We present the numerical results as a
function of $L/\xi$, where $\xi$ is the characteristic length entering
Eq.\ (\ref{eq:Geven}) and Eq.\ (\ref{eq:Godd}). 
We determine $\xi$ by comparing the numerical
data for $L \gg \xi$ to the asymptotic results 
(\ref{eq:Geven}) and (\ref{eq:Godd}).

\begin{figure}
\centerline{\epsfxsize=3in\epsfbox{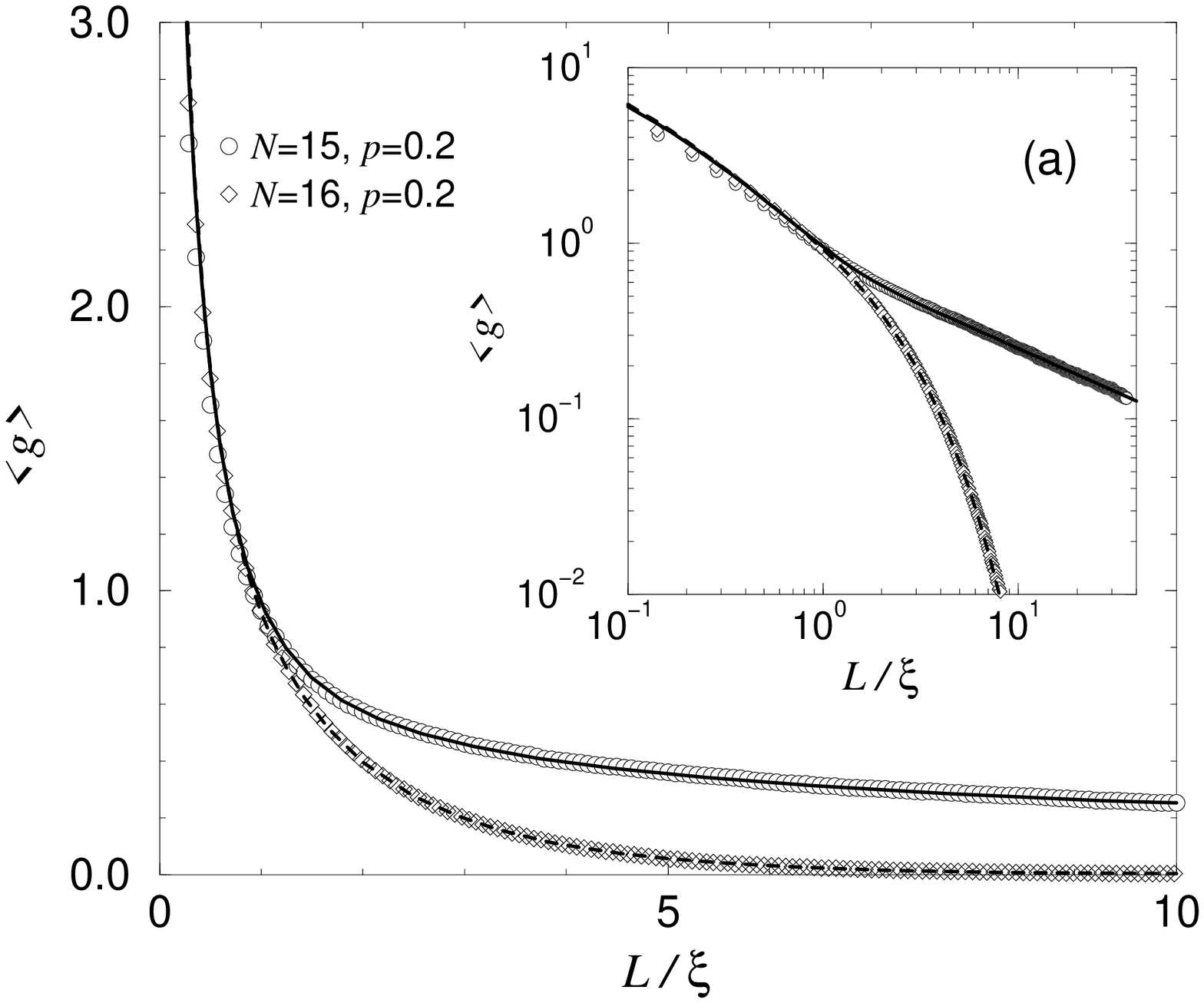}}
\centerline{\epsfxsize=3in\epsfbox{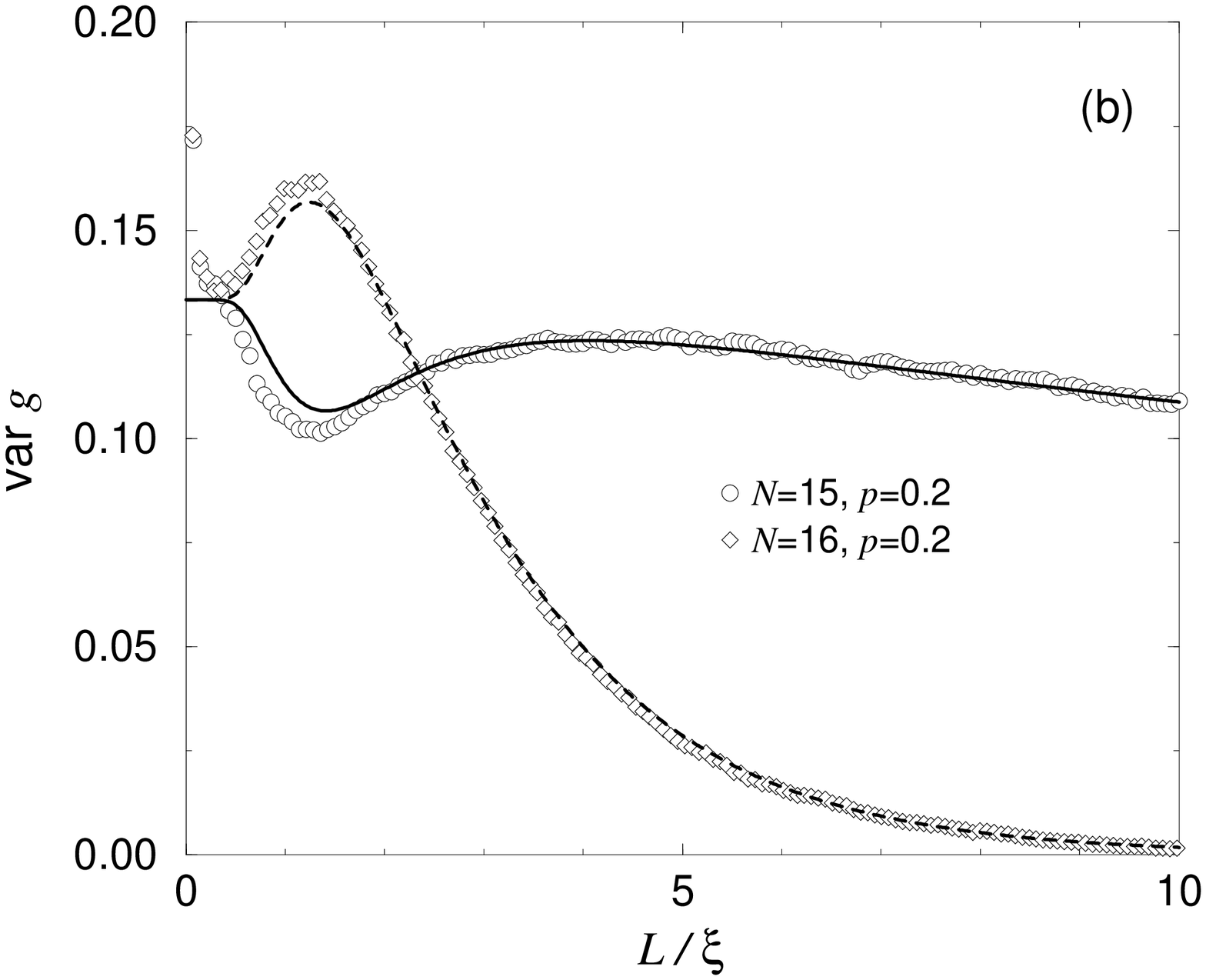}}
\caption{\label{fig: g1516}
The average (a) and the variance (b) of the conductance $g$ for the
random flux model (\protect\ref{eq: def dynamics}) at $\varepsilon = 0$
for $N=15$ (circle) and $N=16$ (diamond) with disorder strength
$p=0.2$. For these parameters, we find that $\xi = 280a$ for $N=15$ and
$\xi = 283a$ for $N=16$.  The solid (dashed) lines in (a) are the
theoretical result (\protect\ref{eq: mean g finite N}) for $\langle g
\rangle$ for $N=15$ ($16$); the solid (dashed) lines in (b) are the
large odd (even) $N$ analytical results (\protect\ref{eq: var g, large
N}) for $\mbox{var}\, g$.}
\end{figure}

Figure \ref{fig: g1516} shows the average and variance of the 
conductance at $\varepsilon=0$
of the random flux model (\ref{eq: def dynamics}) with $N=15$ and
$N=16$ and disorder strength $p=0.2$.  When $L\gg\xi$, $\langle g
\rangle$ decreases algebraically for $N=15$ whereas it decays
exponentially for $N=16$.  This is precisely the even-odd
effect\cite{Miller96,Brouwer98} that we discussed at length in the
last section.  We find excellent agreement between the numerical data
and the theory of Sec.\ \ref{sec:3}, which is indicated by the solid
(odd $N$) and dashed (even $N$) lines in the figure.  
The characteristic length $\xi$ 
that governs the crossover to the slow algebraic decay of Eq. (\ref{eq:Godd})
is estimated to be $280a$ for $N=15$.
The localization length $\xi$ 
that governs the exponential decay of Eq. (\ref{eq:Geven})
is estimated to be $283a$ for $N=16$.
As in the case of the average conductance,
for $\mbox{var}\, g$, the even-odd effect can be clearly seen for
$L\gtrsim\xi$, where the numerical data coincide with the analytic
result in the large $N$ limit, Eq.~(\ref{eq: var g, large N}).
The slight discrepancy at very small $L$ happens at $M \sim N$ and may be
understood as a crossover from quasi one-dimensional to
quasi two-dimensional behavior.
This type of one- to two-dimensional crossover was reported in a
numerical work by Tamura and Ando.\cite{TamuraAndo}
The Fokker-Planck approach employed in sections 
\ref{sec:The random magnetic flux model} 
and 
\ref{sec:Moments of the conductance}
is specifically
devised for a quasi one-dimensional geometry and is therefore
inapplicable to describe the regime $M\lesssim N$.

\begin{figure}
\centerline{\epsfxsize=3in\epsfbox{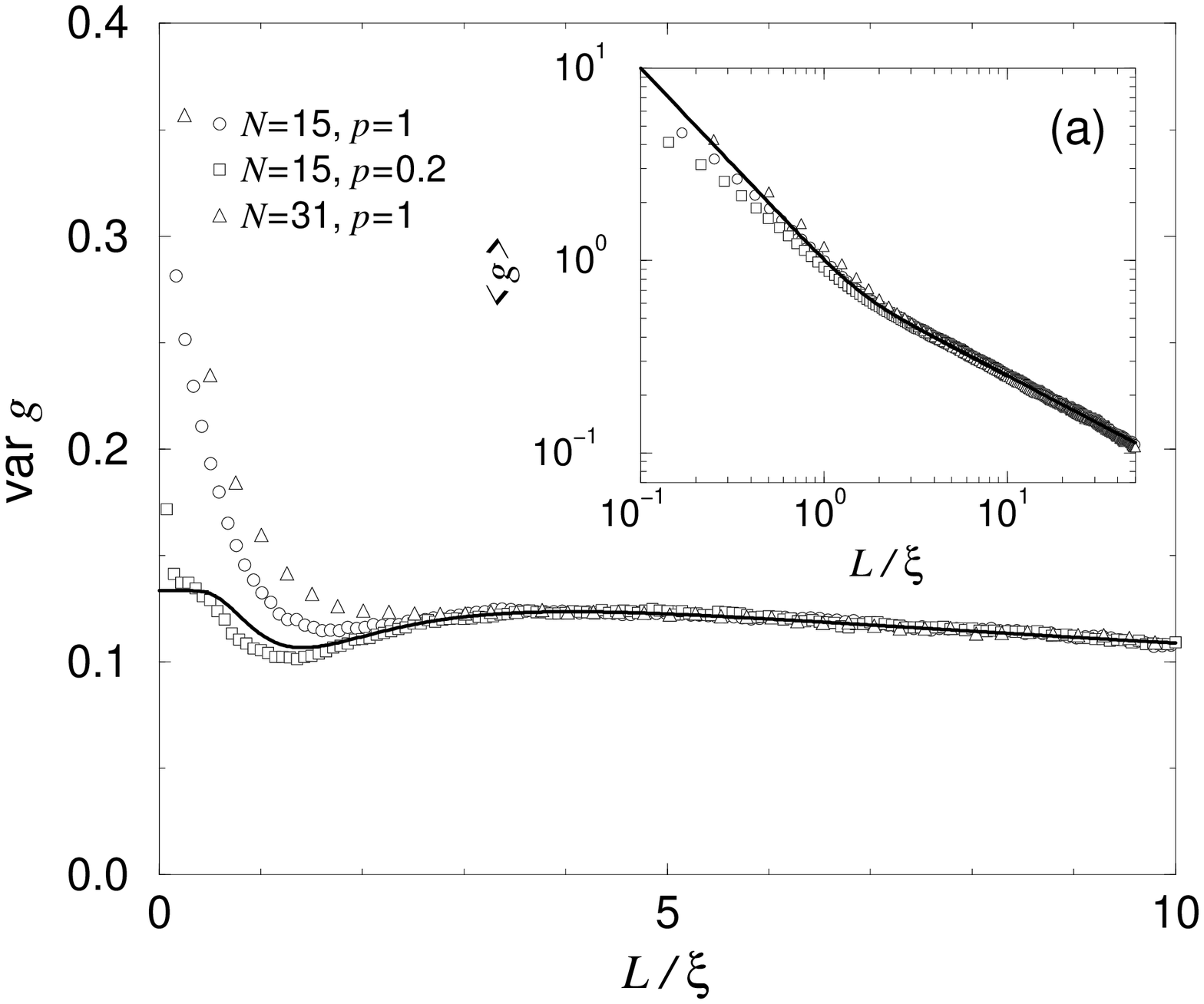}}
\centerline{\epsfxsize=3in\epsfbox{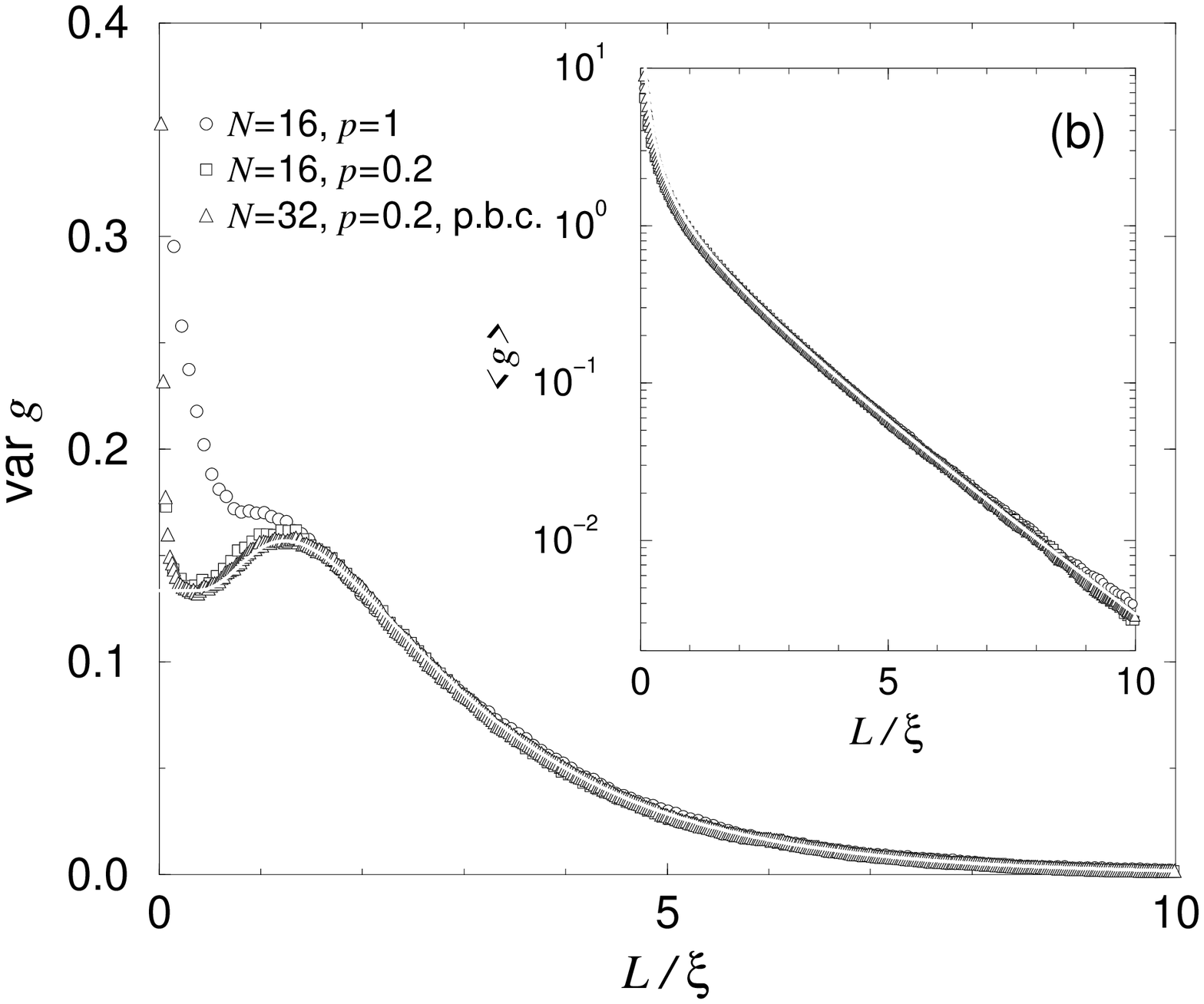}}
\smallskip
\caption{\label{fig: oddN}
The average and the variance of the conductance $g$ for the
random flux model (\protect\ref{eq: def dynamics}) at $\varepsilon=0$
for odd $N$ (a) and even $N$ (b).  
(a) 
The circles, squares, and triangles are the results for
$(N,p)=(15,1.0)$, $(15,0.2)$, and $(31,1.0)$, respectively.  The
characteristic length $\xi$ is numerically found to be $\xi=23.7a$,
$280a$, and $39.8a$, respectively.  The solid lines are the large odd
$N$ analytical result (\protect\ref{eq: var g, large N}).
(b)
The circles, squares, and triangles are the results for $(N,p)=(16,1.0)$,
$(16,0.2)$, and $(32,0.2)$, respectively. In the last case, periodic
boundary conditions (p.b.c.) in the transverse directions are used,
whereas for the first two cases open boundary conditions are assumed.
The localization lengths $\xi$ are found to be $\xi=27.2a$, $283a$,
and $476a$, respectively.  The white solid lines are the large even $N$
analytical result for the chiral-unitary class of section
\protect\ref{sec:Moments of the conductance}.
}
\end{figure}

\begin{figure}
\centerline{\epsfxsize=3in\epsfbox{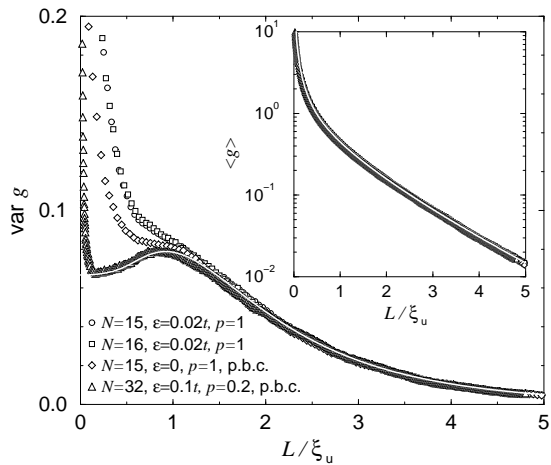}}
\smallskip
\caption{\label{fig: unitary}
Average and variance of $g$ for disorder strength $p=1$ away from the
critical energy $\varepsilon=0$.  The circles and squares are the data
for $N=15$ and $\varepsilon=0.02t$ ($\xi_{\rm u}=47.8a$) and for $N=16$ and
$\varepsilon=0.02t$ ($\xi_{\rm u}=50.6a$), respectively.  The diamonds and
triangles are the data for $N=15$ and $\varepsilon=0$ ($\xi_{\rm u}=45.0a$) 
and for $N=32$ and $\varepsilon=0.1t$ ($\xi_{\rm u}=1041a$), 
which are calculated for the periodic boundary condition. 
For odd $N$, periodic boundary conditions break the chiral symmetry.  
The white solid lines are the large $N$
analytical result for the unitary class, taken from
Refs.\ \protect\onlinecite{Zirnbauer92,Mirlin94,Frahm95}.
}
\end{figure}

In Figs.\ \ref{fig: oddN} and \ref{fig: unitary} we consider the
dependence of $\langle g \rangle$ and $\mbox{var}\, g$ on $N$, $p$, and
$\varepsilon$. Figure \ref{fig: oddN}(a) shows the average and variance of
$g$ for odd $N$ at $\varepsilon = 0$ for two choices of $N$ and $p$.
We see that the numerical data show fairly good agreement with the
analytic large $N$ result (solid lines) for the three cases we
examined. For larger disorder strength $p$, the deviations from the
analytical result (\ref{eq: var g, large N}) is more prominent, the
stronger disorder data being closer to the onset of quasi
two-dimensional behavior
for small $L$. The agreement between the numerical data for $p=1$ and
the theory of Sec.\ \ref{sec:3} for $L \gg \xi$ is remarkable, in view
of the fact that the theory was derived under the assumption of weak
disorder, whereas $p=1$ corresponds to the strongest possible disorder
in the random flux model.

Figure \ref{fig: oddN}(b) shows $\langle g \rangle$ and $\mbox{var}\, g$
for the random flux model (\ref{eq: def dynamics}) at $\varepsilon=0$
for an even number $N$ of channels.  We show the data for three cases
$(N,p)=(16,1)$, (16,0.2), and (32,0.2).  In the last example we used
periodic boundary conditions in the transverse direction instead of the
open boundary conditions of Eq.\ (\ref{eq: def dynamics}). Since $N$ is
even, the periodic boundary conditions do not destroy the chiral
symmetry, so that the system remains in the chiral unitary symmetry
class.  We see that the results of numerical simulations are
indistinguishable from the theoretical curves (solid lines) for both
$\langle g\rangle$ and ${\rm var}\ g$ except in the quasi
two-dimensional regime $M\lesssim N$.  
We conclude from Fig.\ \ref{fig: oddN}(a) and \ref{fig: oddN}(b)
that the localization properties of the random flux model
at $\varepsilon=0$ are governed by the chiral unitary
universality class, independent of the disorder strength.

In figure \ref{fig: unitary} we show some results where the chiral
symmetry is broken. In this case, charge transport is no longer
governed by the Fokker-Planck equation (\ref{eq:DMPKChiral}) for the
chiral unitary symmetry class, but by the Fokker-Planck equation
(\ref{eq:DMPK}) that is valid for the standard unitary
class.\cite{Dorokhov,MPK,BeenakkerReview} In the figure, numerical
results are shown for three cases away from the critical energy as well
as for one case where the chiral symmetry does not exist because of the
periodic boundary condition imposed for odd $N$.  With the exception of
very short lengths, where the system becomes quasi two-dimensional, all
the data for $\langle g \rangle$ and for $\mbox{var}\, g$ agree with
the theoretical prediction for the unitary
class.\cite{Zirnbauer92,Mirlin94,Frahm95} The results indicate that the
small nonzero energy $\varepsilon=0.02t$ is sufficient to cause a
change from the chiral unitary symmetry class at $\varepsilon=0$ to the
standard unitary symmetry class.  Another interesting feature to note
is that the localization length $\xi$ in Fig.~\ref{fig: unitary} is
roughly twice as large as that in the chiral case (Fig.\ \ref{fig: oddN}). 
(For example, compare the two cases
$\varepsilon = 0$ and $\varepsilon = 0.02t$ for $N=16$ and $p=1$, where
we find $\xi = 27.2a$ and $\xi_{\rm u} = 50.6a$, respectively.) 
This behavior was observed earlier in Refs.\
\onlinecite{Sugiyama2,Miller96,Furusaki98}.  This result is consistent
with the analytic result that $\xi$ differs by a factor of 2 between
the chiral ($\xi=N\ell$) and the unitary class ($\xi_{\rm u}=2N\ell$),
assuming that the mean free path determined by the short-distance
physics is identical in the two classes.  (For the numerical results we
may expect that the mean free path should not have strong energy
dependence on the scale of $|\varepsilon|<0.1t$.)

\section{Conclusion}
\label{sec:Conclusions}

In this paper we studied transport properties of a particle on a
rectangular lattice in the presence of uncorrelated random fluxes of
vanishing mean. This problem is commonly known as the random flux
problem. We considered a wire geometry and weak disorder 
and showed that the symmetries
of the random flux problem have dramatic consequences on the
statistical distribution of the conductance $g$. If the energy
$\varepsilon$ is away from the band center $\varepsilon=0$, the system
belongs to the standard unitary symmetry class, while at
$\varepsilon=0$, transport is governed by an additional symmetry of the
random flux model, the particle-hole or chiral symmetry. 
We have compared numerical simulations of the average and variance of the
conductance $g$ in the random flux model in a thick quantum wire 
to analytical calculations for
the standard unitary and the chiral unitary symmetry classes, and found
good agreement for $\varepsilon \neq 0$ and $\varepsilon=0$, respectively.

There are important differences between the conductance distribution in
the chiral unitary symmetry class and the standard unitary symmetry
class, both in the diffusive and the localized regime. These
differences are summarized in Table \ref{tab:1}. The most striking
feature of the chiral unitary symmetry class is the even-odd
effect:\cite{Miller96,Brouwer98} If the number of channels $N$ in the
wire is even, the average conductance $\langle g \rangle$ decays
exponentially with length $L$ in the localized regime $L \gg N \ell$,
whereas for odd $N$, the decay of $\langle g \rangle$ is algebraic.
The sensitivity to the chiral symmetry in transport properties is very
strong.  For example, removing the chiral symmetry by a change in
boundary condition is sufficient to change the universality class to
the standard unitary one, even in the thick-wire limit $N\gg1$.

\begin{table}
\begin{tabular}{l|c|c|c}
 & unitary & \multicolumn{2}{c}{chiral unitary} \\
 & & even $N$ & odd $N$ \\ 
\hline
 diffusive &&& \\
 $\langle g \rangle$ & ${N^{\vphantom{M}} \ell \over L_{\vphantom{M}}}$ & 
${N \ell \over L}$ & ${N \ell \over L}$ \\
 $\mbox{var}\, g$ & ${1^{\vphantom{M}} \over 15_{\vphantom{M}}}$ & 
${2 \over 15}$ & ${2 \over 15}$ \\ 
\hline
 localized & $\vphantom{ M^M \over M_M}$ && \\
 $\langle g \rangle$ & $\vphantom{ M^M \over M_M} 
2\left({\pi N \ell \over L} \right)^{3/2} e^{-L/4 N \ell}$ & 
$\sqrt{{8 N \ell \over \pi L}} e^{-L/2 N \ell}$ 
 & $\sqrt{{2 N \ell \over \pi L}}$ \\
 $\mbox{var}\, g$ &  $\vphantom{ M^M \over M_M} {1 \over 2} 
\left({\pi N \ell \over L} \right)^{3/2} e^{-L/4 N \ell}$ & 
$\sqrt{{2 N \ell \over \pi L}} e^{-L/2 N \ell}$ 
 & $\sqrt{{8 N \ell \over 9 \pi L}}$ \\
\end{tabular}\hfill\\

\caption{\label{tab:1} Average and variance of the conductance $g$ of a
disordered quantum wire with $N \gg 1$ channels and mean free path
$\ell$ in the unitary and chiral unitary symmetry classes, for the
diffusive regime $L \ll N \ell$ and for the localized regime $L \gg N
\ell$. The results for the unitary ensemble are taken from Refs.\ \protect\onlinecite{Zirnbauer92,Mirlin94,Frahm95}.}
\end{table}

Although our theory is limited to a quasi one-dimensional geometry and
cannot account for the crossover from one to two dimensions, it does
show the importance of the chiral symmetry to the transport properties
of the random flux problem. Taking the prominent role played by
symmetry for the random flux model in quasi one-dimension as a
guideline, we speculate that a similar picture is appropriate for
the two-dimensional random flux problem.
Hence, following Gade and Wegner,\cite{Gade93} and  Miller and 
Wang\cite{Miller96} we expect that the
localization properties of the two-dimensional random flux problem are
controlled by the unitary symmetry class away from the band center
$\varepsilon=0$, whereas the band center $\varepsilon=0$ plays the role
of a critical energy.
The random flux problem would thus share with the Integer Quantum Hall
Effect, and with the problem of Dirac fermions in a random vector
potential the existence of a single critical energy that lies between
energies with localized states.
There are however two important differences with the Integer Quantum
Hall Effect. First, there is no symmetry that fixes the value of the
critical energy in the Integer Quantum Hall Effect, while the chiral
symmetry of the random flux model implies that criticality
occurs at the band center $\varepsilon=0$.  Second, in contrast to the
smooth density of states in the Integer Quantum Hall Effect,
one expects that the density of states in the random flux problem
is singular at $\varepsilon=0$. Such a singularity
of the density of states at $\varepsilon = 0$ was observed in the
single chain random hopping problem, is suggested by
the numerical simulations
of Refs.\ \onlinecite{Minakuchi96,Furusaki98}, 
and is consistent with Gade's analysis of the two-dimensional
non-linear-$\sigma$ model with chiral symmetry,\cite{Gade93} 
and with  exact results on the problem of
Dirac fermions in a random vector potential.\cite{Ludwig94,Nersesyan94} 
(The latter problem
shares the same chiral symmetry as the random flux problem although it
differs from the random flux problem in that the magnetic fluxes are
strongly correlated on all length scales.)

\acknowledgments

We are indebted to A.\ Altland, K.\ M.\ Frahm, 
B.\ I.\ Halperin, D.\ K.\ K.\ Lee, P.\ A.\ Lee, M. Sigrist, N.\ Taniguchi and 
X.-G.\ Wen for useful discussions.
AF and CM would like to thank P.\ A.\ Lee and R.\ Morf for their kind
hospitality at MIT and PSI, respectively, where parts of this work were
completed.
PWB acknowledges support by the NSF under grant nos.\ DMR 94-16910, DMR
96-30064, and DMR 97-14725.  
CM acknowledges a fellowship from the Swiss Nationalfonds.
AF is supported by a Monbusho grant for overseas research and is 
grateful to the condensed-matter group at Stanford University for
hospitality. 
The numerical calculations were performed on workstations at the Yukawa
Institute, Kyoto University.

\appendix

\section{Derivation of the Fokker-Planck equation}
\label{ap: Derivation of FP}

In this paper, we described the transport properties of a quantum wire
in the chiral unitary symmetry class in terms of its transfer matrix ${\cal
M}$. Our theoretical analysis was focused on a solution of the
Fokker-Planck equation (\ref{eq:DMPKChiral}) that governs the
$L$-evolution of the probability distribution $P(x_1,\ldots,x_N;L)$ of
the eigenvalues $e^{\pm 2 x_j}$ of ${\cal M} {\cal M}^{\dagger}$. A
derivation of this Fokker-Planck equation from a different microscopic
model was presented in Ref.\ \onlinecite{Brouwer98}. Here we present an
alternative derivation of Eq.\ (\ref{eq:DMPKChiral}) that is closer
in spirit to derivations of the Fokker-Planck equation for the unitary
class existing in the literature.\cite{MPK}

For the statistical distribution of the parameters $x_j$, the
symmetries of the transfer matrix ${\cal M}(\varepsilon)$ are of
fundamental importance. For the random flux model, there are
two symmetries (cf.\ Sec.\ \ref{sec:The random magnetic flux model}):
\begin{eqnarray}
  {\cal M}(\varepsilon)\,\Sigma_3\,{\cal M}^{\dagger}(\varepsilon)
  = \Sigma_3,\ \ && \mbox{flux conservation}, 
\label{eq:fluxcons} \\
  \Sigma_1 {\cal M}(\varepsilon) \Sigma_1 = {\cal M}(-\varepsilon),
  \ \ && \mbox{chiral symmetry}. 
\label{eq:chiralsym}
\end{eqnarray}
Here the transfer matrix ${\cal M}$ is defined in Eq.\ (\ref{eq:Mdef})
and $\Sigma_j = \sigma_j \otimes \openone_N$, where $\sigma_j$ is the
Pauli matrix ($j=1,3$) and $\openone_N$ is the $N \times N$ unit
matrix.

Because of flux conservation (\ref{eq:fluxcons})
${\cal M}(\varepsilon)$ can be parameterized as\cite{BeenakkerReview}
\begin{eqnarray}
{\cal M}&=& \pmatrix{{\cal M}_{11} & {\cal M}_{12} \cr {\cal M}_{21} &{\cal M}_{22}\cr} \nonumber \\ &=&
\pmatrix{{\cal U}&0\cr 0&{\cal U}'\cr}
\pmatrix{\cosh X&\sinh X\cr \sinh X&\cosh X\cr}
\pmatrix{{\cal V}&0\cr 0&{\cal V}'\cr},
\label{eq:cal M if no chiral} \label{eq:cal M chiral}
\end{eqnarray}
where ${\cal U}$, ${\cal U}'$, ${\cal V}$, and ${\cal V}'$ are $N
\times N$ unitary matrices and $X$ is a diagonal matrix containing the
parameters $x_j$ on the diagonal. We are interested in the case of zero
energy, when the chiral symmetry (\ref{eq:chiralsym}) results in the
further constraints ${\cal U} = {\cal U}'$ and ${\cal V} = {\cal V}'$.
Notice that in this case, with the parameterization (\ref{eq:cal M if
no chiral}), the parameters $x_j$ are uniquely determined by ${\cal M}$.
This is an important difference with the unitary symmetry class, where
each $x_j$ is only defined up to a sign. As a result, in the unitary
class, the distribution $P(x_1,\ldots,x_N;L)$ has to be symmetric under
a transformation $x_j \to -x_j$ for each $j$ individually, while no
such symmetry requirement exists in the chiral unitary class.\cite{symmetry}


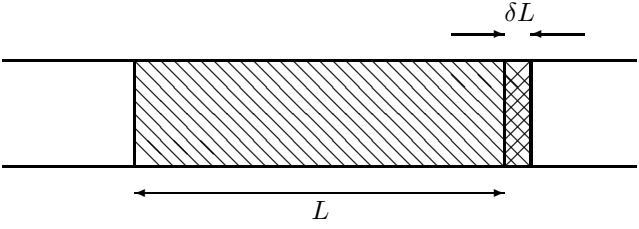
\begin{figure}
\begin{picture}(240,110)(0,-100)
%
%
%
%
%
%
%
%
%
\thicklines
\put(  0,-10){\line(1,0){240}}
\put(  0,-50){\line(1,0){240}}
%
%
\put( 50,-50){\line(0,1){40}}
\put(190,-50){\line(0,1){40}}
\thinlines
\put(120,-60){\vector(-1,0){70}}
\put(120,-60){\vector( 1,0){70}}
\put(117,-70){$L$}
\thicklines
\put(200,-50){\line(0,1){40}}
\thinlines
\put(170,  0){\vector( 1,0){20}}
\put(220,  0){\vector(-1,0){20}}
\put(190,  5){$\delta L$}
\put(50,-15){\line(1,-1){35}}
\put(50,-20){\line(1,-1){30}}
\put(50,-25){\line(1,-1){25}}
\put(50,-30){\line(1,-1){20}}
\put(50,-35){\line(1,-1){15}}
\put(50,-40){\line(1,-1){10}}
\put(50,-45){\line(1,-1){ 5}}
\multiput(50,-10)(5,0){23}{\line(1,-1){40}} 
\put(200,-15){\line(-1,1){ 5}}
\put(200,-20){\line(-1,1){10}}
\put(200,-25){\line(-1,1){15}}
\put(200,-30){\line(-1,1){20}}
\put(200,-35){\line(-1,1){25}}
\put(200,-40){\line(-1,1){30}}
\put(200,-45){\line(-1,1){35}}
\multiput(190,-50)(0,5){7}{\line(1,1){10}} 
\end{picture}
\caption
{
A thin slice of length $\delta L$ with $a\ll\delta L\ll\ell\ll L$
is added to the disordered region of length $L$.
}
\label{fig:wire}
\end{figure}

As the length $L$ of the disordered region is increased
(see Fig. \ref{fig:wire}), the parameters
$x_j$, $j=1,\ldots,N$ are subjected to a Brownian motion process: As $L$ is
increased by an amount $\delta L$, the parameters $x_j$ will undergo a
(random) shift $x_j \to x_j + \delta x_j$. We first seek the appropriate
Langevin equations that describe the statistical distribution of the
increments $\delta x_j$. Hereto we note that the transfer matrix $\hat
{\cal M} \equiv {\cal M}(0;L+\delta L)$ is the product of the
individual transfer matrices ${\cal M} \equiv {\cal M}(0;L)$ and
${\cal M}' \equiv {\cal M}(0; \delta L)$ for wires of length 
$L$ and $\delta L$, respectively:
\begin{equation}
  \hat {\cal M} = 
     {\cal M} {\cal M}'.
\end{equation}
We also use that the matrix
\begin{equation}
  2 {\cal M}_{11} {\cal M}_{12}^{\dagger} =
  {\cal U}\, \sinh (2 X)\, {\cal U}^{\dagger}
\end{equation}
is hermitian and has eigenvalues $\sinh 2 x_j$, $j=1,\ldots,N$. Hence
we find that
\begin{eqnarray}
 2 \hat {\cal M}_{11} \hat {\cal M}_{12}^{\dagger} &=&
  {\cal U} \left( \sinh2X + 2 \Delta \right) {\cal U}^{\dagger},
  \nonumber \\ \Delta &=&
  {\cal U}^{\dag}\left(
  \hat {\cal M}^{\ }_{11}\, \hat {\cal M}^{\dag}_{12}\,-\,
       {\cal M}^{\ }_{11}\,      {\cal M}^{\dag}_{12} \right){\cal U}.
\end{eqnarray}
Making use of the symmetries of ${\cal M}'$ and of the parameterization
(\ref{eq:cal M chiral}) we can rewrite $\Delta$ as
\begin{eqnarray}
  \Delta &=& 
  \cosh X\, {\cal V} {\cal M}_{12}' {\cal M}_{12}'^{\dagger} 
    {\cal V}^{\dagger}\,  \sinh X
  \nonumber \\ && \mbox{} +
  \sinh X\, {\cal V} {\cal M}_{12}' {\cal M}_{12}'^{\dagger} 
    {\cal V}^{\dagger}\,  \cosh X
  \nonumber \\ && \mbox{} +
  \cosh X\, {\cal V} {\cal M}_{11}' {\cal M}_{12}'^{\dagger} 
    {\cal V}^{\dagger}\,  \cosh X
  \nonumber \\ && \mbox{} +
  \sinh X\, {\cal V} {\cal M}_{12}' {\cal M}_{11}'^{\dagger} 
    {\cal V}^{\dagger}\,  \sinh X.
\end{eqnarray}
We take the length $\delta L$ of the added slice small compared to the
mean free path $\ell$. Within the thin slice the disorder is assumed to be
uncorrelated beyond a length scale of the order of the lattice spacing
$a\ll \delta L$.
In this case one has ${\cal M}' = 1 + {\cal
O}(\delta L)^{1/2}$, so that the matrix $\Delta$ is of order $(\delta
L)^{1/2}$ itself and we can treat it in perturbation theory.  As a
result, we find that the addition of the slice of width $\delta L$
results in the change
\begin{eqnarray}
\sinh 2 \hat x_j- \sinh 2 x_j
  &=& 2 \Delta_{jj}+ 4
\sum_{k\neq j}
{\Delta_{jk} \Delta_{kj} \over
\sinh 2 x_j-\sinh 2 x_k} 
\nonumber \\ && \mbox{} 
 + {\cal O}(\delta L^{3/2}),
\end{eqnarray}
%
%
or equivalently
\begin{eqnarray}
  \delta x_j & = & 
{\Delta_{jj}\over \cosh 2 x_j} - 
{\Delta_{jj}^2 \sinh 2 x_j \over \cosh^3 2 x_j} 
\nonumber \\ 
&& \mbox{} + 2
\sum_{k\neq j}
{\Delta_{jk} \Delta_{kj} \over
(\sinh 2 x_j - \sinh 2 x_k)\cosh2 x_j} 
\nonumber \\ 
&& \mbox{}+{\cal O}(\delta L^{3/2}). \label{eq:deltax}
\end{eqnarray}

It remains to find the first two moments of $\Delta_{jk}$. 
Hereto we make an ansatz for the distribution of the transfer
matrix ${\cal M}'$. Because ${\cal M}'$ is close to $1$, it is natural
to parameterize it in terms of its generator,
\begin{equation}
  {\cal M}' = \exp\, {\cal A}.
\end{equation}
{}From the symmetry requirements (\ref{eq:fluxcons}) and
(\ref{eq:chiralsym}) we deduce that ${\cal A}$ has the form
\begin{equation}
  {\cal A} = i V \otimes \openone_2 + W \otimes \sigma_1,
\end{equation}
where $V$ and $W$ are hermitian $N \times N$ matrices. 
We choose a convenient statistical distribution of ${\cal M}'$ 
by assuming that $V$ and $W$ have independent, 
Gaussian distributions with zero mean and with variance
\begin{eqnarray}
  \langle V_{ij} V_{kl} \rangle = \langle W_{ij} W_{kl} \rangle = 
 \delta_{il} \delta_{jk} {\delta L \over N \ell}.
\end{eqnarray}
Then we find that the first two moments of $\Delta$ are given by
\begin{eqnarray*}
\langle \Delta_{jk} \rangle &=&
\delta_{jk} \sinh (2 x_j) {\delta L \over \ell}, \\
\langle \Delta_{jk} \Delta_{kj} \rangle &=&
\cosh^2(x_j + x_k) {\delta L \over N \ell}. \\
\end{eqnarray*}
Combining this with Eq.\ (\ref{eq:deltax}), we conclude that under addition of a
narrow slice of width $\delta L \ll \ell$, the parameters $x_j$ undergo a
shift $x_j \to x_j + \delta x_j$ with
\begin{mathletters}
\label{eq: moments of increments}
\begin{eqnarray}
&&
\langle\delta x_j\rangle_{\delta L} =
{\delta L\over N \ell}\sum_{k \neq j}\coth( x_j- x_k), 
\label{eq:n=1}\\
&&
\langle\delta x_j \delta x_k\rangle_{\delta L}=
{\delta L\over N \ell}\delta_{jk},
\label{eq:n=2}
\end{eqnarray}
\end{mathletters}%
all higher moments vanishing to first order in $\delta L$.
Equation (\ref{eq: moments of increments}) is equivalent to
the Fokker-Planck equation (\ref{eq:DMPKChiral}).

\section{Solution to the Fokker-Planck equation}
\label{ap: P({x_j})}

In this appendix, we present an exact solution for the Fokker-Planck
equation (\ref{eq:DMPKChiral}), closely following the exact solution of
the DMPK equation in the unitary symmetry class by Beenakker and
Rejaei.\cite{BeenakkerRejaei} We start by rewriting
Eq.\ (\ref{eq:DMPKChiral}) as
\begin{mathletters} \label{eq:FP with Omega}
\begin{eqnarray}
\ell {\partial P\over\partial L} &=&
{1\over2 N }\sum_{j=1}^N{\partial\over\partial x_j}
\left[
{\partial P\over\partial x_j}+
2 P\left({\partial\Omega\over\partial x_j}\right)
\right],
\\
\Omega &=&-{1\over 2}
\sum_{j<k} \ln|\sinh(x_j-x_k)|^2,
\label{eq:Omega in FP}
\end{eqnarray}
where the initial condition is
\begin{eqnarray}
P(x_1,\ldots,x_N;0)=\prod_{j=1}^N\delta(x_j).
\label{eq:FP initial cond}
\end{eqnarray}
\end{mathletters}

The key step towards the exact solution of Eq.\ (\ref{eq:FP with Omega})
is the transformation
\begin{equation}
P(\{x_j\};L)=
\left[\prod_{j<k}\sinh(x_j-x_k)\right]  \Psi(\{x_j\};L),
\label{eq: trsf with Omega}
\end{equation}
which changes the Fokker-Planck equation (\ref{eq:FP with Omega}) into
a Schr\"odinger equation,
\begin{eqnarray}
- \ell {\partial\Psi\over\partial L} &=&
-{1\over2 N }\sum_{j=1}^N{\partial^2\Psi\over\partial x^2_j}
+{1 \over 2 N}\Psi\,\sum_{j=1}^N
\left[
\left({\partial\Omega\over\partial x_j}\right)^2-
{\partial^2\Omega\over\partial x^2_j}
\right] \nonumber \\ &=&
-{1\over2 N }\sum_{j=1}^N{\partial^2\Psi\over\partial x^2_j}
+ U \Phi. \label{eq:Schrod}
\end{eqnarray}
Here $U = (N-1) (N-2)/6 + (N-1)/2$.  Thus, $\Psi(x_1,\ldots,x_N;L)$
obeys a Schr\" odinger equation in imaginary time
$L$ that describes $N$ identical free particles on the line,
$-\infty<x<\infty$. (For comparison, in the unitary symmetry class, one
finds that $\Psi$ obeys a Schr\"odinger equation for $N$ identical
particles moving in the presence of a potential $\propto \sinh^{-2} 2x$
which repels the $x$'s away from the origin.\cite{BeenakkerRejaei})
 
Since the probability distribution $P(x_1,\ldots,x_N;L)$ is
symmetric under a permutation of the $x_j$'s, it follows from Eq.
(\ref{eq: trsf with Omega}) that $\Psi(x_1,\ldots,x_N;L)$ must be
antisymmetric, i.e., it must describe the imaginary-time evolution of
$N$ identical fermions.
At $L=0$, the initial condition (\ref{eq:FP initial cond}) implies that
all $x_j$ coincide at the origin.  Hence, at $L=0$, the transformation
(\ref{eq: trsf with Omega}) is singular. We avoid this problem by
starting with the initial condition\cite{BeenakkerReview} 
\begin{eqnarray}
P(\{x_j\};0|\{y_k\}) &=& {1\over N!}
\sum_{\sigma}\prod_{j=1}^N\delta(x_j-y_{\sigma(j)}),
\label{eq:new FP initial cond} \nonumber \\ y_j = \epsilon (j-1),
\end{eqnarray}
where all the initial values are different, and send $\epsilon$ to zero
at the end of the calculation. 
The summation is over all permutations $\sigma$ of $1,\ldots,N$.

To solve Eq.\ (\ref{eq:Schrod}), we denote by $G(x;L|y)$ the
single-particle Green function of the diffusion
equation obeying
\begin{equation}
\ell {\partial G\over\partial L}=
{1\over2 N}\, {\partial^2 G\over\partial x^2},
\qquad G(x;0|y)=\delta(x-y). \label{eq:Geq}
\end{equation}
Solution of Eq.\ (\ref{eq:Geq}) yields 
\begin{equation}
G(x;L|y)= \sqrt{N \ell \over 2 \pi L}\, e^{-{N \ell \over 2 L}(x-y)^2}.
\end{equation}
Then the Slater determinant
\begin{eqnarray}
\Psi(\{x_j\};L|\{y_k\})&=&
{1\over N!}\, {\rm det}\, \left[G(x_j;L|y_k)\right]_{j,k=1,\ldots,N}
\nonumber\\&&\mbox{}\times
e^{-UL/\ell} 
\label{eq: slater determinant}
\end{eqnarray}
is antisymmetric in $x_1,\ldots,x_N$ and obeys the Schr\"odinger
equation (\ref{eq:Schrod}). Using the inverse of the transformation
(\ref{eq: trsf with Omega}), we obtain that
\begin{eqnarray}
&&
P(\{x_j\};L|\{y_k\})=
\nonumber\\
&&
\Psi(\{x_j\};L|\{y_k\})
\prod_{j<k}{\sinh(x_j-x_k)\over\sinh(y_j-y_k)}
\label{eq: sol FP with x|y}
\end{eqnarray}
is the solution to the Fokker-Planck equation (\ref{eq:FP with Omega})
with the regularized initial condition (\ref{eq:new FP initial cond}).

We finally take the limit $\epsilon \to 0$. This limit must be treated
with care in view of the denominator of
Eq.\ (\ref{eq: sol FP with x|y}).  With the help of
\begin{eqnarray}
&&
{\rm det}\, \left[e^{-{N\ell\over2L}(x_j-y_k)^2}\right]_{j,k=1,\ldots,N}=
\\
&&
e^{-\sum\limits_{j=1}^N {N \ell \over 2 L} x_j^2}
\left({N \ell \epsilon\over 2L}\right)^{N(N-1)\over2}
\prod_{j<k}(x_j-x_k) +{\cal O}(\epsilon^2),
\nonumber
\end{eqnarray}
the singularity $\propto \epsilon^{-N(N-1)/2}$ 
coming from the denominator in Eq.\ (\ref{eq: sol FP with x|y}) is cancelled. 
We thus recover Eq.\ (\ref{eq: jpd if unitary}).



\begin{references}
\bibitem{footnote} 
The mean free path $\ell$ in Eq.\ (\ref{eq:DMPKChiral}) differs a factor $2$
from the one used in Ref.\ \onlinecite{Brouwer98}. The choice made in
Eq.\ (\ref{eq:DMPKChiral}) ensures that in the diffusive regime $\ell \ll
L \ll N \ell$ the average conductance is given by the Drude formula
$\langle g \rangle = N \ell/L$.
\bibitem{symmetry} 
An equivalent way to deal with the non-uniqueness of the parameters
$x_j$ appearing in the parameterization (\ref{eq:cal M if no chiral})
for the unitary class is to require that all $x_j$ are positive. In
that picture, the $x_j$ are positive for the unitary class, while
they can be both positive and negative for the chiral unitary class.
\end{references}
\end{document}